\documentclass[a4paper]{amsart}

\usepackage{style}
\begin{document}

\title{Index Calculus in the Trace Zero Variety}

\author{Elisa Gorla}
\address{Elisa Gorla, Institut de math\'ematiques, Universit\'e de Neuch\^atel, Rue Emile-Argand 11, 2000 Neuch\^atel, Switzerland}
\email{\href{mailto:elisa.gorla@unine.ch}{elisa.gorla@unine.ch}}
%\thanks{The authors were supported by the Swiss National Science Foundation under grant no.\ 123393.}
%\affil{\small Institut de math\'ematiques, Universit\'e de Neuch\^atel, Rue Emile-Argand 11, 2000 Neuch\^atel, Switzerland, \small {\tt elisa.gorla@unine.ch}}

\author{Maike Massierer}%\samethanks[1]}
\address{Maike Massierer, Mathematisches Institut, Universit\"at Basel, Rheinsprung 21, 4051 Basel, Switzerland}
\email{\href{mailto:maike.massierer@unibas.ch}{maike.massierer@unibas.ch}}
%\affil{\small Mathematisches Institut, Universit\"at Basel, Rheinsprung 21, 4051 Basel, Switzerland, {\tt maike.massierer@unibas.ch}}

%14G50 Applications (of Arithmetic problems/Diophantine geometry) to coding theory and cryptography
%11G25 Varieties over �nite and local �elds
%14H52 Elliptic curves over global �elds
%11T71 Algebraic coding theory; cryptography
%14K15 Arithmetic ground �elds
%11Y40 Algebraic number theory computations

\begin{abstract}
We discuss how to apply Gaudry's index calculus algorithm for abelian varieties to solve the discrete logarithm problem in the trace zero variety of an elliptic curve. We treat in particular the practically relevant cases of field extensions of degree 3 or 5. Our theoretical analysis is compared to other algorithms present in the literature, and is complemented by results from a prototype implementation.
\end{abstract}

\maketitle

\section{Introduction} \label{sec:intro}

Given an elliptic curve $E$ defined over a finite field $\Fq$, consider the group $E(\Fqn)$ of rational points over a field extension of prime degree $n$. Since $E$ is defined over $\Fq$, the group $E(\Fqn)$ contains the subgroup $E(\Fq)$ of $\Fq$-rational points of $E$. Moreover, it contains the subgroup $T_n$ of points $P\in E(\Fqn)$ whose trace $P+\varphi(P)+\ldots+\varphi^{n-1}(P)$ is zero, where $\varphi$ denotes the Frobenius homomorphism on $E$. The group $T_n$ is called the trace zero subgroup of $E(\Fqn)$, and it is the group of $\Fq$-rational points of the trace zero variety relative to the field extension $\Fqn|\Fq$. 

In this paper, we study the hardness of the DLP in the trace zero variety. Our interest in this question has several motivations. First of all, supersingular trace zero varieties can achieve higher security per bit than supersingular elliptic curves, as shown by Rubin and Silverberg in~\cite{rubin-silverberg-02,rubin-silverberg-09} and by Avanzi and Cesena in \cite{avanzi-cesena-07,cesena-10}. Ideally, in pairing-based protocols the embedding degree $k$ is such that the DLP in $T_n$ and in $\F_{q^{kn}}^*$ have the same complexity. In order to achieve this, an accurate assessment of the complexity of the DLP in $T_n$ is necessary.
Moreover, since $T_n$ is isomorphic to $E(\Fqn)/E(\Fq)$, the DLP in $E(\Fqn)$ has the same complexity as the DLP in $T_n$. This provides another motivation to study the hardness of the DLP in $T_n$. A further motivation comes from the fact that the trace zero subgroup itself can be used within asymmetric cryptographic protocols using the DLP as a primitive, as proposed by Frey in~\cite{frey-98}.

Using trace zero varieties in cryptographic protocols presents some advantages with respect to elliptic curves. In fact, a clever use of the Frobenius endomorphism allows us to compute the group operation more efficiently than for an elliptic curve of about the same cardinality, leading to more efficient scalar multiplication in the group (see~\cite{frey-99, lange-phd, lange-04, avanzi-cesena-07} or \cite[Section~15.3.2]{handbook-hecc}). This technique is analogous to the one for Koblitz curves \cite{koblitz-91} and was later applied to GLV--GLS curves \cite{glv,gls}. Another advantage is that for groups of cryptographically relevant size, the order of the group can simply be calculated using the characteristic polynomial of the Frobenius endomorphism. This allows for more efficient computation of the group order in comparison to the group of rational points of an elliptic curve over a prime field of comparable size  (see~\cite[Section~15.3.1]{handbook-hecc}).
Finally, in the recent papers~\cite{gorla-massierer-1} and~\cite{gorla-massierer-2} we proposed new efficient representations for the elements of $T_n$, for any $n$. More precisely, we can represent the elements of the group with $(n-1)\log_2 q+1$ bits, which is optimal since $|T_n|\sim q^{n-1}$. We refer the interested reader to~\cite{gorla-11} for a discussion of the relevance of efficient representations.

In this paper, we discuss how to apply Gaudry's index calculus algorithm for abelian varieties to solve the discrete logarithm problem in $T_n$. Gaudry's algorithm first appeared in~\cite{gaudry-09}, and proposes a general framework to do index calculus on a general abelian variety. A main difficulty of running an index calculus attack on an abelian variety is producing the relations. When the abelian variety is an elliptic curve, Gaudry proposes to use Semaev polynomials (\cite{semaev-04}) to build a system of polynomial equations, such that a solution to the system corresponds to a relation.
The systems can be solved by Gr\"obner bases methods. The complexity of this attack depends on the size of $\Fq$ and the dimension of the abelian variety: Asymptotically in $q$, and regarding $n$ as a constant, it has complexity $\tilde{O}(q^{2-2/(n-1)})$, which is lower than that of generic attacks on $T_n$ and on $E(\Fqn)$ for $n\geq 5$. This leads to the lowest-complexity attack on the DLP in $E(\Fqn)$ for prime~$n$. Other attacks, of comparable or lower complexity but which only apply to specific elliptic curves, are discussed in~\cite{ghs, diem-ghs, diem-06, diem-kochinke, diem-scholten}. We apply Gaudry's index calculus algorithm to $T_n$, and demonstrate that it is feasible for $n=3$ and $q$ up to about 30 bits. % (then $E(\F_{q^3})$ is a 90-bit group). 
%Our computations show that, when $n=3$, the index calculus attack is faster than a Pollard--Rho attack on $E(\F_{q^3})$ for $\log_2 q \geq 30$ approximately. 
For $n=5$ we show that the bottleneck of the algorithm is the Gr\"obner basis computation. Using some tricks from~\cite{bettale-faugere-perret-08,joux-vitse-12} we are able to produce relations and  to solve a DLP for very small $q$, but the attack this yields is not feasible over fields of cryptographic size, therefore it is presently not a threat to the DLP in $T_5$ or $E(\F_{q^5})$. 

We also analyze the algorithm asymptotically in $n$ and $q$, and we see that the complexity is exponential in $n$. This is mostly due to the fact that in order to produce relations, the algorithm solves polynomial systems whose size (number of equations, number of indeterminates, degrees of the equations) depends on $n$, and that the Gr\"obner basis methods have a large complexity in these parameters. We conclude that one can only hope to produce relations with this method for small values of $n$.

The paper is organized as follows. We recall the functionality of index calculus algorithms and the most important definitions in connection with the trace zero variety in Section \ref{sec:prelim}. Then we describe the application of Gaudry's algorithm to the trace zero variety in Section \ref{sec:ictzv}, and we analyze its complexity in Section \ref{sec:complexity}. In Section \ref{sec:experimentsindexcalc}, we present explicit equations and Magma experiments for $n=3,5$. Finally, we compare the index calculus attack with other attacks on the DLP in $T_n$ in Section \ref{sec:comparison}, and discuss the implications of our results for trace zero elliptic curve cryptosystems in Section \ref{sec:securitydiscussion}.

\subsubsection*{Acknowledgements} We thank Pierrick Gaudry, Peter Schwabe, Vanessa Vitse, and Bo-Yin Yang for useful discussions on the material of this paper. We are grateful to the mathematics department of the University of Z\"urich for access to their computing facilities. The authors were supported by the Swiss National Science Foundation under grant no.\ 123393.

\section{Preliminaries} \label{sec:prelim}

\subsection{Index calculus}\label{sec:gaudrysattack}

The security of several public key cryptosystems, including ElGamal and DSA, is based on the hardness of the discrete logarithm problem.

\begin{definition}\label{def:dlp}
Let $G$ be a finite additive group. Given two elements $P \in G$ and $Q \in \langle P \rangle$,
the {\it discrete logarithm problem (DLP)} is
$$ \text{find an element } \ell \in \Z/(\ord P) \Z \text{ such that } \ell P = Q. $$
The number $\ell$ is called the {\it discrete logarithm of $Q$ in base $P$}, and denoted by $\log_P Q$.
\end{definition}

A combination of the Pollard--Rho Algorithm %\cite{pollard-78} 
and the Pohlig--Hellman Algorithm %\cite{pohlig-hellman-78} 
can solve an instance of the DLP in any group $G$ of known order $|G|$ in time $\tilde O(\sqrt{p})$, where $p$ is the largest prime factor of $|G|$.

However, when a concrete group is given, its properties can often be exploited in order to devise more efficient attacks. A particularly powerful such class of attacks are {\it index calculus algorithms}, which exploit the algebraic structure of the groups that they work in. There are index calculus algorithms that compute the DLP in multiplicative groups of finite fields (namely the number field sieve for prime fields \cite{adleman-79,gordon-93,joux-lercier-03} and the function field sieve for fields of small to medium characteristic \cite{coppersmith-84-2,adleman-94,adleman-demarrais-huang-94,schirokauer-02,joux-lercier-02,joux-lercier-06,joux-12,gologlu-granger-mcguire-zumbragel-13-1,joux-13,gologlu-granger-mcguire-zumbragel-13-2,caramel-13,barbulescu-gaudry-joux-thome-13}), elliptic curves over extension fields \cite{semaev-04,gaudry-09,diem-11-2,diem-11-3}, Picard groups of hyperelliptic curves and more generally $C_{a,b}$ curves \cite{adleman-demarrais-huang-94,gaudry-00,enge-02,diem-06,diem-thome-08,enge-gaudry-07,enge-08,enge-gaudry-thome-11,velichka-jacobson-stein-11}, and even general abelian varieties \cite{gaudry-09}.

The general outline of an index calculus attack is as follows (see e.g.\ \cite{enge-gaudry-02}). Let us assume that the goal is to compute a discrete logarithm $\ell = \log_P Q$ of an element $Q \in \langle P \rangle$ in some group $G$. Since we are only working in the cyclic subgroup, we may assume that  $G = \langle P \rangle$. %, and we let $N := |G| = \ord(P)$.
\begin{enumerate}[1.]
 \item {\bf Factor base:} Choose a factor base $\fb = \{P_1,\ldots,P_k\} \subseteq \langle P \rangle$.
 \item {\bf Relation collection:} Construct relations of the form
 $ \alpha_j P + \beta_j Q = \sum_{i=1}^k m_{ij}P_i $
 for $j = 1,\ldots,r > k$.
 \item {\bf Linear algebra:} Given the matrix $M = (m_{ij}) \in (\Z/\ord(P)\Z)^{k \times r}$, compute a non-zero column vector $\gamma = (\gamma_1,\ldots,\gamma_r)^{\intercal}$ in the right kernel of $M$.
 \item {\bf Individual logarithm:} Output $\ell = -(\sum_{j=1}^r \alpha_j \gamma_j)(\sum_{j=1}^r \beta_j \gamma_j)^{-1}$ if $\sum \beta_j \gamma_j$ is invertible in $\Z/\ord(P)\Z$, otherwise return to step 2.
\end{enumerate}

It is easy to see that this gives the correct result: Since $\gamma$ is in the right kernel of $M$, we have $M \gamma = 0$, or equivalently
$$ \sum_{j=1}^r m_{ij}\gamma_j = 0 \quad \text{ for all } i = 1,\ldots,k. $$
Multiplying all relations from step 2 by $\gamma_j$, summing over $j$, and using the above equality gives
$$ \sum_{j=1}^r \alpha_j\gamma_j P + \sum_{j=1}^r \beta_j\gamma_j Q = \sum_{j=1}^r \sum_{i=1}^k m_{ij}\gamma_j P_i = \sum_{i=1}^k \left(\sum_{j=1}^r m_{ij} \gamma_j \right) P_i = 0. $$
Therefore,
$$ Q = -\left(\sum_{j=1}^r \alpha_j\gamma_j\right)\left(\sum_{j=1}^r \beta_j\gamma_j\right)^{-1}P = \ell P. $$

Algorithms that function in this way have been used for many years to compute discrete logarithms in groups where a concept of factorization is available. However, it was not until 2009 that Gaudry \cite{gaudry-09} published an algorithm that works in abelian varieties of dimension at least 2. His idea is to translate the condition for a relation into a system of polynomial equations and to solve the system with Gr\"obner basis methods in order to obtain relations. We give more details on his approach in Section \ref{sec:ictzv}, where we apply it to the trace zero variety. The heuristic complexity of his attack is $\tilde{O}(q^{2-2/d})$ asymptotically for $q \rightarrow \infty$, where the dimension $d \geq 2$ and all other parameters associated to the variety (like the degrees of the defining equations and the size of the representation) are assumed to be constant or bounded by constants.

Since its publication, Gaudry's algorithm has been applied mostly to the Weil restriction of elliptic curves defined over extension fields. In fact, Gaudry suggests this application himself in his original article \cite{gaudry-09}. A similar algorithm for elliptic curves was developed independently by Diem \cite{diem-11-2}. The algorithm of Gaudry and Diem was implemented by Joux and Vitse \cite{joux-vitse-12}. With several further improvements and variations, including a specialized implementation of the Gr\"obner basis algorithm F4 \cite{joux-vitse-11} using an idea of Traverso \cite{traverso-89}, they were able to solve an instance of an oracle-assisted static Diffie--Hellman problem in $E(\F_{2^{155}})$, which is related to, but easier than, the DLP in the same group \cite{granger-joux-vitse-10}. Faug\`ere, Perret, Petit, and Renault \cite{faugere-perret-petit-renault-12}, Petit and Quisquater \cite{petit-quisquater-12}, and Shantz and Teske \cite{shantz-teske-13} studied the polynomial systems that arise during this attack. They come to the conclusion that these systems are of a special shape and that special-purpose Gr\"obner basis techniques may lead to a significant speed-up. The application of the algorithm to Edwards curves was studied by Faug\`ere, Gaudry, Huot, and Renault in \cite{huot-12,huot-13}. 

Notice that this approach only threatens elliptic curves defined over extension fields and does not affect groups $E(\F_p)$ where $p$ is a prime. The best attack on such groups is the Pollard--Rho attack, and the current record for computing a discrete logarithm in $E(\F_p)$, for $p$ a 112-bit prime, is held by Bos, Kaihara, Kleinjung, Lenstra, and Montgomery \cite{ec-record}, using a parallelized version of the Pollard--Rho Algorithm. Some improvements which take into account the use of the negation map in running the Pollard--Rho Algorithm are discussed in~\cite{bls-11}.

Besides elliptic curves, Gaudry's algorithm for abelian varieties has been applied to the Weil restriction of hyperelliptic curves of small genus by Nagao \cite{nagao-07} and to algebraic tori by Granger and Vercauteren \cite{granger-vercauteren-05}. In this paper, we apply Gaudry's attack to the trace zero variety.

\subsection{The Trace Zero Variety}\label{sec:prelimtzv}

Throughout this paper, let $E$ be a smooth elliptic curve defined over a finite field $\Fq$ by an affine Weierstra\ss\ equation. For any extension field $\F$ of $\Fq$, the $\F$-rational points $E(\F)$ on $E$ form a group with neutral element $\O$, the point at infinity. When $\F = \Fqn, n \geq 1,$ is a finite extension, $E(\Fqn)$ is a finite group of order about $q^n$. We denote by $+$ the group operation and by $\varphi$ the Frobenius endomorphism on $E$
$$ \varphi : E \rightarrow E, \quad (X,Y) \mapsto (X^q,Y^q), \quad \O \mapsto \O.$$
Throughout the paper, we denote field elements by uppercase and indeterminates by lowercase letters.

\begin{definition}
 For a field extension $\Fqn|\Fq$ of degree $n > 1$, the {\it trace map} is defined by
 $$ \Tr : E(\Fqn) \rightarrow E(\Fq), \quad P \mapsto P + \varphi(P) + \ldots + \varphi^{n-1}(P). $$
 When $n$ is prime, the kernel of the trace map is called the {\it trace zero subgroup} of $E(\Fqn)$. We denote it by $T_n$.
\end{definition}

The trace zero subgroup is isomorphic to the group of $\Fq$-rational points of the {\it trace zero variety} $V_n$, which is an $(n-1)$-dimensional subvariety of the Weil restriction of $E$: Fixing a basis $\{\zeta_0,\ldots,\zeta_{n-1}\}$ of $\Fqn|\Fq$, % and the Weil restriction coordinates
%\begin{eqnarray*}
% x & = & x_0\zeta_0+\ldots+x_{n-1}\zeta_{n-1}\\
 %y & = & y_0\zeta_0+\ldots+y_{n-1}\zeta_{n-1},
%\end{eqnarray*}
we have $V_n(\Fq) \cong T_n$ via
\begin{equation}\label{eqn:iso}
 (X_0,\ldots,X_{n-1},Y_0,\ldots,Y_{n-1}) \mapsto (X_0\zeta_0+\ldots+X_{n-1}\zeta_{n-1},Y_0\zeta_0+\ldots+Y_{n-1}\zeta_{n-1}). 
\end{equation}
In this paper, we consider the case $n \geq 3$, when the trace zero variety has dimension at least 2.

We study the hardness of the DLP in $T_n$, which is of interest in cryptography for various reasons, as explained in the Introduction. In particular, the DLP in $T_n$ is as hard as the DLP in $E(\Fqn)$. This is shown for the analogous case of algebraic tori in \cite{granger-vercauteren-05}, and more generally for exact sequences of abelian varieties in \cite{galbraith-smith-06}. The result as we state it here is Proposition~2.4 in~\cite{gorla-massierer-2}.

%Scalar multiplication of points in $T_n$ is particularly efficient, since it can be sped up using the Frobenius endomorphism, see \cite{frey-99, lange-phd, lange-04, avanzi-cesena-07}. This technique is analogous to the one for Koblitz curves \cite{koblitz-91} and was later applied to GLV--GLS curves \cite{glv,gls}. 
%Due to the efficient arithmetic, trace zero subgroups were proposed for the use in public-key cryptosystems by Frey \cite{frey-99}. 
%In this paper, we study the hardness of the DLP in trace zero subgroups.

%Trace zero subgroups are also interesting in the context of pairing-based cryptography, where they achieve the largest security parameters in some cases \cite{rubin-silverberg-02,rubin-silverberg-09,avanzi-cesena-07,cesena-10}.

%Moreover, the DLP in $T_n$ is as hard as the DLP in $E(\Fqn)$. This is shown for the analogous case of algebraic tori in \cite{granger-vercauteren-05}, and more generally for exact sequences of abelian varieties in \cite{galbraith-smith-06}. The result as we state it here is Proposition~2.4 in~\cite{gorla-massierer-2}.

\begin{proposition}
 Let $E$ be an elliptic curve defined over $\Fq$, and let $T_n$ be the trace zero subgroup of $E(\Fqn)$ for some prime number $n$. Then the sequence
 $$ 0 \longrightarrow E(\Fq) \longrightarrow E(\Fqn) \overset{\varphi - \id}{\longrightarrow} T_n \longrightarrow 0 $$
 is exact, and the DLP in $E(\Fqn)$ has the same complexity as the DLP in $T_n$.
\end{proposition}

In \cite{gorla-massierer-1} we wrote an equation for the $x$-coordinates of the points in $T_n$ using the Semaev polynomial. We briefly summarize how to write such an equation, starting with the definition and the main result from \cite{semaev-04}.

\begin{definition}
 Let $\Fq$ be a finite field of characteristic at least 5, and let $E$ be a smooth elliptic curve defined over $\Fq$ by the affine equation
 $$ E: y^2 = x^3 + Ax + B. $$
 The $m$-th {\it summation polynomial} or {\it Semaev polynomial} $f_n$ is defined recursively by
$$\begin{array}{rcl}
f_3(z_1,z_2,z_3) & = & (z_1 - z_2)^2z_3^2 - 2((z_1+z_2)(z_1z_2 + A) + 2B)z_3 + (z_1z_2-A)^2 - 4B(z_1+z_2) \\
f_m(z_1,\ldots,z_m) & = & \Res_z(f_{m-k}(z_1,\ldots,z_{m-k-1},z),f_{k+2}(z_{m-k},\ldots,z_m,z)) 
\end{array} $$
for $m \geq 4$ and $m-3 \geq k \geq 1$, where $\Res$ denotes the resultant.
\end{definition}

\begin{theorem}[\cite{semaev-04}, Theorem 1]\label{thm:semaev}
For any $m \geq 3$, let $Z_1,\ldots, Z_m$ be elements of the algebraic closure $\Fqbar$ of $\Fq$. Then $f_m(Z_1,\ldots,Z_m) = 0$ if and only if there exist $Y_1,\ldots,Y_m \in \Fqbar$ such that the points $(Z_i,Y_i)$ are on $E$ and $(Z_1,Y_1) + \ldots + (Z_m,Y_m) = \O$ in the group $E(\Fqbar)$. Furthermore, $f_m$ is absolutely irreducible and symmetric of degree $2^{m-2}$ in each variable. The total degree is $(m-1)2^{m-2}$.
\end{theorem}

\begin{remark}\label{rmk:char2ic}
The original definition from \cite{semaev-04} is for elliptic curves defined over fields of characteristic at least 5. However, polynomials with the same properties can be defined also for characteristic 2 and 3. Therefore, all results of this paper hold, with the appropriate adjustments, over finite fields of any characteristic.
\end{remark}

The Semaev polynomial is used in \cite{gorla-massierer-1} to give the following equation for the $x$-coordinates of the points of $T_n$.

\begin{proposition}[{\cite[Proposition 3, Remark 5]{gorla-massierer-1}}]\label{thm:exceptions}
 Let $n$ be an odd prime, and let $T_n$ be the trace zero subgroup of $E(\Fqn)$. Then
 $$ T_n \subseteq \{ (X,Y) \in E(\Fqn) \mid f_n(X,X^q,\ldots,X^{q^{n-1}}) = 0 \} \cup \{\O\}. $$
 Moreover, we have
 $$\begin{array}{lcl}
    T_3 & = & \{ (X,Y) \in E(\F_{q^3}) \mid f_3(X,X^q,X^{q^2}) = 0\} \cup \{\O\}  \\
    T_5 \cup (E[3](\F_q)+(E[2]\cap T_5)) & = & \{ (X,Y) \in E(\F_{q^5}) \mid f_5(X,X^q,\ldots,X^{q^4}) = 0 \} \cup \{\O\}.
 \end{array}$$
In the case when $n=3$ or 5, for any root $X \in \Fqn$ of $f_n(x,x^q,\ldots,x^{q^{n-1}}) = 0$ it can be decided efficiently whether $(X,Y) \in T_n$ by checking $Y \in \Fqn$ and, if $n=5$, by checking in addition that $X\notin \mathcal{L} := \{ X_{Q+R}\mid Q+R=(X_{Q+R},Y_{Q+R}) \in E[3](\Fq)+(E[2]\cap T_5), \; Q \ne \O \}$, where $|\mathcal{L}| \leq 16$.
\end{proposition}

As discussed in \cite{gorla-massierer-1} at the end of Section~3, Weil restriction of $f_n(x,x^q,\ldots,x^{q^{n-1}})=0$ with respect to the coordinates
\begin{eqnarray*}
 x & = & x_0\zeta_0+\ldots+x_{n-1}\zeta_{n-1}\\
 y & = & y_0\zeta_0+\ldots+y_{n-1}\zeta_{n-1}
\end{eqnarray*}
and reduction modulo the polynomials $x_i^q-x_i$ yield exactly one equation
\begin{equation}\label{eqn:semaevic}
\tilde{f}_n(x_0,\ldots,x_{n-1}) = 0. 
\end{equation}
Its zeros describe the $x$-coordinates of the points of $V_n(\Fq)$ as given by Proposition \ref{thm:exceptions} and via the isomorphism (\ref{eqn:iso}). %The polynomial has total degree $(n-1)2^{n-2}$.
Therefore, we henceforth use (\ref{eqn:semaevic}) as an equation for the trace zero subgroup. It has total degree $(n-1)2^{n-2}$.

\section{An index calculus algorithm for the trace zero variety}\label{sec:ictzv}

Following the ideas of Gaudry \cite{gaudry-09}, we propose the following index calculus algorithm to compute discrete logarithms in $T_n$. When $n=2$, then $V_n$ is one-dimensional, and the attack cannot be applied. Therefore, we only consider $n \geq 3$. Furthermore, we assume that $T_n$ is cyclic, which is the most relevant case in cryptography. %For given $q$ and $n$, curves $E$ such that $T_n$ is cyclic are easy to find, in fact in our experiments we did not encounter a single pair $(q,n)$ where we could not find a suitable elliptic curve $E$.

\begin{remark}\label{rmk:notcyclic}
 When $T_n$ is not cyclic, some of the probability estimates in Section \ref{sec:complexity} may be wrong and the algorithm may not function as expected. However, these problems can be overcome using classical randomization techniques (see \cite[Remark 2]{gaudry-09}, \cite{enge-gaudry-02}).
\end{remark}

The algorithm takes as input two points $P,Q \in T_n$ such that $T_n = \langle P \rangle$, and it outputs the discrete logarithm $\log_P Q$, i.e.\ a number $\ell = \log_P Q \in \Z/\ord(P)\Z$ such that $\ell P = Q$ in $T_n$. Below, we describe the different steps of the algorithm in detail. We always identify $T_n$ and $V_n(\Fq)$ via the isomorphism (\ref{eqn:iso}).

\subsection{Setup}
Following the suggestion of Semaev \cite{semaev-04}, we carry out the index calculus algorithm working only with the $x$-coordinates of points in $T_n$. We choose a basis $\{\zeta_0,\ldots,\zeta_{n-1}\}$ of the extension $\Fqn|\Fq$ and represent an affine point $P = (X,Y) \in T_n$ via the coordinates
$$ P = (X_0,\ldots,X_{n-1}), $$
where $X = X_0\zeta_0 + X_1 \zeta_1 + \ldots + X_{n-1}\zeta_{n-1}$. So by writing $(X_0,\ldots,X_{n-1}) \in T_n$ we mean that there exists a $Y$ such that $(X,Y) \in T_n$. 
%As we saw in Chapter \ref{sec:semaeveqn}, all such points satisfy a certain equation. 
We use (\ref{eqn:semaevic}) as an equation for $T_n$.

%As in Section \ref{sec:prelimtzv}, denote by $f_n(z_1,\ldots,z_n)$ the $n$-th Semaev polynomial. Then the Weil restriction of $f_n(x,x^q,\ldots,x^{q^{n-1}})$, 
%using $ x = x_0\zeta_0 + x_1 \zeta_1 + \ldots + x_{n-1}\zeta_{n-1} $ and reducing modulo $x_i^q-x_i$ for all $i$,
%yields one single equation, which we denote by $\tilde{f}_n(x_0,\ldots,x_{n-1})$. All points $P \in T_n$ satisfy $\tilde{f}_n(X_0,\ldots,X_{n-1}) = 0$, and conversely the $\Fq$-solutions $(X_0,\ldots,X_{n-1})$ of the equation
%\begin{equation}\label{eqn:semaevic}
% \tilde{f}_n(x_0,\ldots,x_{n-1}) = 0
%\end{equation}
%yield $x$-coordinates of points in $T_n$ via $(X_0,\ldots,X_{n-1}) \mapsto X = X_0\zeta_0 + X_1 \zeta_1 + \ldots + X_{n-1}\zeta_{n-1}$, provided that the corresponding $y$-coordinates are in $\Fqn$ and up to some exceptions. For $n=3$ there are no exceptions, and for $n=5$ the exceptions are well-understood and their number is very small (see Theorem \ref{thm:exceptions}). 

%\begin{remark}
% when we encounter points that are not representable like this, throw away relation. of course this assumes that $P,Q$ are representable in our coordinates.
% never mind, we don't need that. since we're presenting points via their x-coords, this works for any affine point. there's no need for P,Q or aP+bQ to satisfy the equation.
%\end{remark}

\subsection{Factor base}
We define the factor base
$$ \fb = \{ (0,\ldots,0,X_{n-2},X_{n-1}) \in T_n \}. $$
These are the $\Fq$-rational points of a curve $\mathcal{C}$ in $V_n$ obtained by intersecting $V_n$ with the hyperplanes $\{x_0 = 0\},\ldots,\{x_{n-3} = 0\}$. Since $V_n$ has dimension $n-1$, intersecting with $n-2$ hyperplanes generically gives a curve $\mathcal{C}$. Thus $\fb = \mathcal{C}(\Fq)$ has about $q$ elements by the Theorem of Hasse--Weil, provided that $\mathcal{C}$ is absolutely irreducible.

\begin{remark}\label{rmk:fb}
 Important properties of the factor base are that it has about $q$ elements (this will be used in the complexity analysis, see Section \ref{sec:complexity}) and that its elements can be described via algebraic equations (this will allow us to describe relations via a polynomial system, see Section \ref{sec:relations}). A further very important property is that the factor base must generate a large part of $T_n$, so that many elements of $T_n$ decompose over the factor base. For this reason, the curve $\mathcal{C}$ should not be contained in any proper abelian subvariety of $V_n$. Notice that this can easily be detected in practice, since the algorithm will find practically no relations when $\mathcal{C}$ is an abelian subvariety of $V_n$.
 
 Moreover, the fact that $|\fb| \approx q$ can be proven (with the Theorem of Hasse--Weil) only if we assume that $\mathcal{C}$ is smooth and absolutely irreducible. %Notice, however, that intersecting $V_n$ with any choice of $n-2$ hyperplanes, or equivalently, setting any $n-2$ of the $x_i$ equal to zero, may give a factor base with $q$ elements. 
In practice, if setting $x_0=\ldots=x_{n-3}=0$ does not produce a factor base with the desired properties, we simply make a different choice of hyperplanes. In our exposition we assume that the choice we have made is a good one. This is true in all our experiments.
\end{remark}

Using equation (\ref{eqn:semaevic}), we see that any element $(0,\ldots,0,X_{n-2},X_{n-1})\in\fb$ satisfies the equation
%\begin{eqnarray*}
% \tilde{f}_n(X_0,\ldots,X_{n-1}) & = & 0\\
% X_0 & = & 0\\
% & \vdots &\\
% X_{n-3} & = & 0,
%\end{eqnarray*}
%or, by plugging the last equations into the first one,
$ \tilde{f}_n(0,\ldots,0,X_{n-2},X_{n-1}) = 0.$
Conversely, the $\Fq$-solutions $(X_{n-2},X_{n-1})$ of 
\begin{equation}\label{eqn:fbeqn}
 \tilde{f}_n(0,\ldots,0,x_{n-2},x_{n-1}) = 0
\end{equation}
yield $x$-coordinates of points in $\fb$ via $(X_{n-2},X_{n-1}) \mapsto X_{n-2}\zeta_{n-2}+X_{n-1}\zeta_{n-1}$, provided that the corresponding $y$-coordinates are in $\Fqn$ and up to a few exceptions, as explained above (see also Proposition \ref{thm:exceptions}). Therefore, enumerating the factor base essentially amounts to finding all solutions of (\ref{eqn:fbeqn}).

\subsection{Relation collection}\label{sec:relations}
Since $V_n$ has dimension $n-1$, we search for relations of the form
\begin{equation}\label{relation}
 R = P_0  +  \ldots  +  P_{n-2},
\end{equation}
where $R = \alpha P  +  \beta Q \in T_n$ is given and $P_0,\ldots,P_{n-2} \in \fb$ are to be found. We write $U = U_0\zeta_0+U_1\zeta_1+\ldots +U_{n-1}\zeta_{n-1}$ for the $x$-coordinate of $R$.

Following \cite{semaev-04}, we use the Semaev polynomial to describe a relation. If the points $P_0,\ldots,$ $P_{n-2}$ with $x$-coordinates $X_{P_0},\ldots,X_{P_{n-2}}$ are given, then according to Theorem \ref{thm:semaev} they satisfy (\ref{relation}) if and only if
$ f_n(X_{P_0},\ldots,X_{P_{n-2}},U) = 0.$
Therefore, candidates for $x$-coordinates of the $P_i$ can be found by solving
\begin{equation}\label{eqn:relation}
 f_n(x_{P_0},\ldots,x_{P_{n-2}},U)=0
\end{equation}
for the $x_{P_i}$. We apply Weil restriction to equation (\ref{eqn:relation}) using the coordinates
$$ x_{P_i} = x_{i,0}\zeta_0 + x_{i,1}\zeta_1 + \ldots + x_{i,n-1}\zeta_{n-1} $$
and obtain $n$ equations
\begin{equation}\label{sys:rel}
 F_j(x_{0,0},\ldots,x_{n-2,n-1},U_0,\ldots,U_{n-1}) = 0, \quad j = 0,\ldots,n-1.
\end{equation}
Solving this system over $\Fq$ is equivalent to solving equation (\ref{eqn:relation}) over $\Fqn$, and yields possible $x$-coordinates for the points $P_i$.

In addition to requiring that the $P_i$'s sum to $R$, we must ensure that they belong to the factor base. Therefore, we set $x_{i,0} = \ldots = x_{i,n-3} = 0$ for $i = 0,\ldots,n-1$, and we include an equation of the form (\ref{eqn:fbeqn}) in system (\ref{sys:rel}) for each $P_i$. This means that in order to find a relation, we solve the system
\begin{equation}\label{sys:relations}\small
\begin{array}{rcl}
 F_0(0,\ldots,0,x_{0,n-2},x_{0,n-1},\ldots,0,\ldots,0,x_{n-2,n-2},x_{n-2,n-1},U_0,\ldots,U_{n-1}) & = & 0\\
 & \vdots & \\
 F_{n-1}(0,\ldots,0,x_{0,n-2},x_{0,n-1},\ldots,0,\ldots,0,x_{n-2,n-2},x_{n-2,n-1},U_0,\ldots,U_{n-1}) & = & 0\\
 \tilde{f}_n(0,\ldots,0,x_{0,n-2},x_{0,n-1}) & = & 0\\
 & \vdots & \\
 \tilde{f}_n(0,\ldots,0,x_{n-2,n-2},x_{n-2,n-1}) & = & 0
\end{array}
\end{equation}
over $\Fq$. The system has $2n-1$ equations in $2(n-1)$ indeterminates, two indeterminates for each of the $P_i$'s. The first $n$ equations are the Weil descent of the $n$-th Semaev polynomial, where a constant has been plugged in for the last indeterminate. Therefore, they each have total degree at most $(n-1)2^{n-2}$. They describe the condition that the points $P_i$ sum to $R$. The last $n-1$ equations also have total degree at most $(n-1)2^{n-2}$. They guarantee that the solution points $P_i$ belong to the factor base.

Since the system has more equations than unknowns, one would expect that it generically has no solutions over the algebraic closure and that, when it has solutions, then it is zero-dimensional. This is verified in our experiments.
%As observed by Gaudry \cite{gaudry-09}, the system is generically of dimension zero: Denoting by $S_{n-1}$ the $(n-1)$-th symmetric group, consider the map
%$$ \psi : \mathcal{C}^{n-1}/S_{n-1} \rightarrow V_n,\quad (P_1,\ldots,P_{n-1}) \mapsto P_1+\ldots+P_{n-1}. $$
%Since $\mathcal{C}$ is not contained in a proper abelian subvariety of $V_n$ by assumption, the dimension of the image of $\psi$ in $V_n$ is $n-1$, and hence for a generic point $R$ in $V_n$, the number of preimages under $\psi$ over $\Fqbar$ is finite. Therefore, system (\ref{sys:relations}) is generically of dimension zero.
Then, by the Shape Lemma (see e.g.\ \cite[Theorem 3.7.25]{kreuzer-robbiano-1}), the system may be solved by computing a lexicographic Gr\"obner basis and then finding the $\Fq$-roots of a univariate polynomial. Notice that, in order to find the $\Fq$-roots of a polynomial $f(x)\in\Fq[x]$, one would first find the divisor $g(x)$ of $f(x)$ which is the product of all linear factors of $f(x)$ over $\Fq$, then factor $g(x)$, whose degree equals the number of solutions of the system over $\Fq$. Again, this is the case only after a generic change of coordinates. In the examples we computed however, a change of coordinates was never needed.

Whenever a given point $R$ decomposes over the factor base, i.e.\ when a relation of the form (\ref{relation}) exists, this gives a solution of system (\ref{sys:relations}). The converse, however, is not true. For example, when the solutions of the system give $x$-coordinates where one of the corresponding $y$-coordinates is not in $\Fqn$, then this does not produce a valid relation. 
%Notice that the system will often have no solution, or a solution that does not produce a valid relation. 
In theory, it is also possible that a system produces more than one relation. However, we expect this to be extremely rare, since it would produce a relation among the elements of the factor base. 
In accordance with this intuition, we never encountered a system with more than one solution in our experiments.

%\begin{remark}
% The problem that system (\ref{sys:relations}) has solutions that do not produce valid relations because the corresponding $y$-coordinates are not in $\Fqn$ may be avoided by working with the equations for the full (affine part of) the trace zero variety as given in Chapter \ref{sec:deteqn} instead of the Semaev equation (\ref{eqn:semaevic}). However, this would produce a much larger system, namely instead of each $\tilde{f}_n$ there would be $n+1$ equations, and there would be the additional indeterminates $y_{00},\ldots,y_{n-2,n-1}$. The system would then have a total of $n+(n+1)(n-1)=n^2+n-1$ equations and $(2+n)(n-1)= n^2 + n - 2$ indeterminates, and it would be much harder to solve than system (\ref{sys:relations}), which is already quite hard (we will see that it is manageable only for very small values of $n$). In fact, the equations of system (\ref{sys:relations}) generate the elimination ideal of the ideal generated by the larger system, where the variables $y_{ij}$ are eliminated, and therefore the systems are, in a sense, equivalent.
% 
% When $n=3$, building a system from the equations presented in Chapter \ref{sec:freysequations} is also a possibility. They yield a system of 7 equations of total degree at most 4 in 6 indeterminates, whereas the system (\ref{sys:relations}) we propose has only 5 equations of total degree at most 4 in 4 indeterminates. So this system would have two more equations and two more indeterminates, and our experiments showed that computing a Gr\"obner basis for this system is indeed much slower.
%\end{remark}

\begin{remark}\label{rmk:jouxvitse}
 Joux and Vitse \cite{joux-vitse-12} propose considering relations that involve one factor base point less than suggested by Gaudry, i.e.\ only $n-2$ points in our case. This reduces the probability of finding relations by a factor $q$, but in some cases it can make the difference between a manageable and an unmanageable system. We consider this idea in Section \ref{sec:explicit5ic}.
\end{remark}

Finally, we need to produce more relations than there are factor base elements, i.e.\ about $q$, by solving the system sufficiently many times (see Section \ref{sec:complexity} for an estimate) for different random points $R$. 

\subsection{Linear algebra}
The relation collection phase of the algorithm produces a sparse matrix of size about $q \times q$ with entries 0 or 1. Notice that, while it is theoretically possible to have a row whose entries are positive numbers greater than 1, this should be extremely rare and in fact we never encountered such a relation in our experiments. 
The rows of the matrix correspond to the factor base elements, and the columns correspond to the different relations.
% involving a point $R$, each given by the values $\alpha$ and $\beta$. 
Generically a column has $n-1$ non-zero entries, one for each factor base element that appears in the corresponding relation. Assuming that more relations have been produced than there are factor base elements, the matrix has more columns than rows. Therefore, there exists a non-zero vector in its right kernel. The task of the linear algebra step is to find such a vector, where the computations must be performed not over $\Z$, but modulo the order of $P$ in $T_n$. Standard methods to solve such sparse linear systems are Wiedemann's Algorithm and Lanczos' Algorithm (see \cite{wiedemann-86,lamacchia-odlyzko-90}). 

\begin{remark} \label{rmk:la}
 Since there are efficient and well-studied methods for solving sparse linear systems, we do not treat this step in detail. Notice however that the efficient implementation of the linear algebra step is far from trivial, especially since the algorithms are hard to parallelize. One recent record-breaking implementation on GPUs is presented in \cite{jeljeli-12,jeljeli-14}. Moreover, in practice a filtering step can make a big difference, see e.g.\ \cite{bouvier-12}. This is a preprocessing of the matrix, where duplicate relations are removed, points that appear in only one relation (corresponding to rows with only one nonzero entry) are removed, and excess relations are removed until there are exactly $|\fb| + 1$ of them left. We do not employ such sophisticated techniques in our experiments, since we treat only small examples and our emphasis is on finding relations and not on the linear algebra step.
\end{remark}

\subsection{Individual logarithm}
Once the linear system has been solved, computing the actual discrete logarithm is easy. Denoting by $(\gamma_1,\ldots,\gamma_r)$ the vector in the kernel of the matrix computed in the previous step and by $\alpha_j,\beta_j$ the values of $\alpha,\beta$ corresponding to the $j$-th relation we have 
$$\log_P Q = -\left(\sum_{j=1}^r \gamma_j \alpha_j\right)  \left(\sum_{j=1}^r \gamma_j \beta_j\right)^{-1}, $$
provided that $\sum \gamma_j \beta_j$ is invertible modulo the order or $P$. If not, one must collect more relations in order to produce a different matrix and find a different vector $\gamma$. Notice that $\sum \gamma_j \beta_j$ is invertible with high probability, especially if $\ord(P)$ is prime.

%\begin{remark} \label{rmk:char2ic}
% We point out that this algorithm works for elliptic curves defined over finite fields of any characteristic. Although the original definition of the Semaev polynomial in \cite{semaev-04} is only for curves in short Weierstra\ss~ form, it may easily be adjusted to fields of characteristic 2 and 3 (see also Remark \ref{rmk:evencharsemaev}).
%\end{remark}

\section{Complexity analysis}\label{sec:complexity}

We now analyze the complexity of the index calculus algorithm presented in the previous section. We make the same heuristic assumptions as Gaudry \cite{gaudry-09} and other work based on Gaudry's results, e.g.\ \cite{granger-vercauteren-05,joux-vitse-12}. Our analysis is in $q$ and $n$ and therefore more precise than that of Gaudry, who disregards the dependency on $n$. By disregarding the dependency on $n$ in our analysis, one obtains the result of Gaudry. For simplicity we use the $\tilde{O}$-notation, which ignores logarithmic factors in both $n$ and $q$.

\subsection{Setup} Diem \cite{diem-11-2} shows that the $n$-th Semaev polynomial and its Weil restriction can be computed with a randomized algorithm in expected time polynomial in $\tilde{O}(e^{n^2})$.

\begin{remark}
 We do not have to compute the full Weil restriction of $f_n(x_i,x_i^q,\ldots,x_i^{q^{n-1}})$ or of $f_n(x_{P_0},x_{P_1},\ldots,x_{P_{n-2}},u)$, since we only need to evaluate the polynomials on the $x$-coordinates of points in the factor base. Therefore, when computing the Weil restriction, we work with the coordinates $x_{P_i} = x_{i,n-2}\zeta_{n-2} + x_{i,n-1}\zeta_{n-1}$. In practice, this procedure is much quicker than first computing the usual Weil restriction and then setting $x_{i0} = \ldots = x_{i,n-3} = 0$, and the complexity is lower than the one given in \cite{diem-11-2}. However, since this term will not dominate the final complexity of the index calculus algorithm, the complexity estimate by Diem suffices for our purposes.

We choose to treat $u$, the $x$-coordinate of $R$, as an indeterminate. Then we only have to compute the Weil restriction once to obtain system (\ref{sys:relations}). Each time we plug a value for the $x$-coordinate of $R$ into system (\ref{sys:relations}), we obtain a system which possibly produces a relation. 
\end{remark}

\subsection{Factor base} 

In order to enumerate the factor base, we go through all values $X_{n-2} \in \Fq$, compute the solutions of $\tilde{f}_n(0,\ldots,0,X_{n-2},x_{n-1}) = 0$ over $\Fq$, and check whether the solution gives a point in $T_n$. Since the degree of $\tilde{f}_n$ in $x_{n-1}$ is bounded by $(n-1)2^{n-2}$, computing all solutions takes $\tilde{O}((n-1)2^{n-2})$
 operations in $\Fq$ (see \cite[Corollary 14.16]{gerhard-gathen}). Typically, there are only few solutions. Checking whether the $y$-coordinate corresponding to $X = X_{n-2}\zeta_{n-2} + X_{n-1}\zeta_{n-1}$ is in $\Fqn$ is much cheaper. %, since it only requires computing a square root. 
Altogether, enumerating the factor base costs 
 $$\tilde{O}(q(n-1)2^{n-2} ).$$

\subsection{Relation generation}\label{sec:relationgeneration}

Assuming that most {\it different} unordered $(n-1)$-tuples of factor base elements sum to {\it different} points in $T_n$, then $|\fb|^{n-1}/(n-1)!$ points of $T_n$ decompose over the factor base. Since $T_n$ has about $q^{n-1}$ elements, this means that the probability of a point $R \in T_n$ splitting over the factor base is $1/(n-1)!$. Therefore, in order to generate $q$ relations, we expect to have to try to decompose $q(n-1)!$ points, i.e.\ solve $q(n-1)!$ systems.

In order to solve each system, we follow the approach that is most efficient in practice: We first compute a Gr\"obner basis with respect to the degree reverse lexicographic term order, and we then use a Gr\"obner walk algorithm to convert it to a lexicographic Gr\"obner basis. Afterwards, we factor a univariate polynomial. The complexity of the last step is negligible compared to the first two.

To estimate the complexity of the Gr\"obner basis computation, we use the bound on the complexity of Faug\`ere's F5 algorithm \cite{f5}. We assume that the system is semi-regular, which is true generically. Then according to \cite[Proposition 6]{bardet-faugere-salvy-yang-05}, the complexity of computing a degree reverse lexicographic Gr\"obner basis of our system 
%, which is zero-dimensional and in $2n-2$ variables, 
is
$$ O\left( {\dreg + 2n-2 \choose 2n-2}^{\omega}\right), $$
where $2 \leq \omega \leq 3$ is the linear algebra constant (i.e.\ the exponent in the complexity of matrix multiplication) and $\dreg$ is the degree of regularity of the system (this is also called the regularity index, see \cite[Definition 5.1.8]{kreuzer-robbiano-2}).

We estimate $\dreg$ using a standard bound from commutative algebra
%(see \cite{lazard-83}, this is called {\it Macaulay bound} in \cite{bardet-faugere-salvy-yang-05}):
$$ \dreg \leq (2n-2)((n-1)2^{n-2}-1) + 1 = (2n-2)(n-1)2^{n-2} -2n+3. $$
Hence the complexity of computing a degree reverse lexicographic Gr\"obner basis of our system is 
%bounded by
$$ O\left({(2n-2)(n-1)2^{n-2}+1 \choose 2n-2}^{\omega}\right). $$
Now using the FGLM algorithm \cite{fglm}, we may compute from this basis a lexicographic Gr\"obner basis in
$$ O((2n-2) \cdot D^3), $$
where $D$ is the degree of the ideal generated by the degree reverse lexicographic Gr\"obner basis (i.e.\ the number of solutions counted with multiplicity in $\Fqbar$). Using as a bound on $D$ the product of the degrees of $2n-2$ of the equations of the system, we get $$D\leq ((n-1)2^{n-2})^{2n-2}.$$ Therefore, this is not more expensive than F5.

Taking into account that we have to do this $q(n-1)!$ times, the total cost of the relation collection step is
$$ O\left({(2n-2)(n-1)2^{n-2}+1 \choose 2n-2}^{\omega}(n-1)!q\right). $$

\subsection{Linear algebra}\label{sec:linalg}

Using Lanczos' or Wiedemann's Algorithm, the cost of solving a sparse linear system of size about $q \times q$, where each column has $n-1$ non-zero entries, is
$$ O((n-1)q^2) $$
(see e.g.\ \cite{eberly-kalthofen-97}).

\subsection{Individual logarithm} 

The cost of computing the individual logarithm is negligible compared to the complexities above.

Putting everything together, we get that the algorithm has a total complexity of
$$ \tilde{O}\left({(2n-2)(n-1)2^{n-2}+1 \choose 2n-2}^{\omega}(n-1)!q+(n-1)q^2\right). $$

\subsection{Double large prime variation}\label{sec:primevar}

As suggested by Gaudry, we may use the double large prime variation \cite{theriault-03,gandry-thome-theriault-diem-07} in order to rebalance the complexity of the relation collection and the linear algebra step in $q$. Then one must collect $q^{2-2/(n-1)}$ relations instead of $q$ and solve a linear system of size $q^{1-1/(n-1)} \times q^{1-1/(n-1)}$ instead of $q \times q$. Then the overall cost of the algorithm becomes
$$ \tilde{O}\left({(2n-2)(n-1)2^{n-2}+1 \choose 2n-2}^{\omega}(n-1)!q^{2-2/(n-1)}\right). $$
Hence we have proven the following heuristic result.

\begin{theorem}\label{thm:indexcalccomplexity}
 Let $T_n, n \geq 3,$ be the trace zero subgroup of an elliptic curve. 
Then there exists a probabilistic algorithm that computes discrete logarithms in $T_n$ in heuristic time
 $$ \tilde{O}\left({(2n-2)(n-1)2^{n-2}+1 \choose 2n-2}^{\omega}(n-1)!q^{2-2/(n-1)}\right) $$
where $n$ is constant and $q$ tends to infinity. The constant in the $\tilde{O}$ does not depend on $q$ or $n$.
\end{theorem}

The heuristic nature of Theorem \ref{thm:indexcalccomplexity} is due to the following heuristic assumptions, which are standard assumptions in this context, see e.g.\ \cite{gaudry-09}. First of all, we assume that (after a randomization of coordinates) there is a choice of hyperplanes which, upon intersection with $V_n$, produce an absolutely irreducible smooth curve in $V_n$, whose $\Fq$-rational points define a factor base of cardinality about $q$ (see Remark \ref{rmk:fb}), and so that the sums of $n-1$ factor base points produce about $q^{n-1}/(n-1)!$ different elements of $V_n(\Fq)$. Secondly, we assume that the systems to be solved are either empty or zero-dimensional, and semi-regular. Finally, we assume that (after a randomization of coordinates), $T_n$ is cyclic, as explained in Remark \ref{rmk:notcyclic}.

\begin{remark}
In particular, $q$ must be sufficiently large compared to $n$. More precisely, we need that $$q^{n-1} > (n-1)!q^{2(1-1/(n-1))(1-2/(n-1))}.$$
This is due to the fact that we need to be able to find enough relations: Taking into account the double large prime variation, we need to produce  $q^{2-2/(n-1)}$ relations, and the probability of finding a relation is $1/((n-1)!q^{1-2/(n-1)})$. Therefore we expect to have to try to decompose about $(n-1)!q^{2(1-1/(n-1))(1-2/(n-1))}$ points, hence $T_n$ must have at least that many points.
\end{remark}

If we allow the constant in the $\tilde{O}$ to depend on $n$ but not on $q$, Theorem \ref{thm:indexcalccomplexity} gives the heuristic complexity of $\tilde{O}(q^{2-2/(n-1)})$ from \cite{gaudry-09}. Our analysis makes the exponential dependency on $n$ explicit. The exponential dependency of the complexity on $n$ was already pointed out by Gaudry and is due to the cost of the Gr\"obner basis computation. Notice that one cannot hope to get subexponential complexity in $n$ for generic systems, due to the complexity bound for F5, which is exponential in $n$ in our situation.
%There are instances of Gr\"obner basis computations which are doubly exponential in the degrees of the input equations (see e.g.~\cite{mayr-meyer-82}).

\section{Explicit equations and experiments}\label{sec:experimentsindexcalc}

We now study the systems of polynomial equations that describe the relations and the overall behavior of our algorithm for $n=3,5$. All computations were done with Magma version 2.19.3 \cite{magma} on one core of an Intel Xeon X7550 Processor (2.00 GHz) on a Fujitsu Primergy RX900S1. Our implementation is only meant to be a proof of concept. It is a straightforward implementation of the algorithm described in Section \ref{sec:ictzv}, and we use the built-in Magma routines wherever possible, e.g.\ for Gr\"obner basis computation, polynomial factorization, and linear algebra. Our timings are only meant as an indication, and they could be improved significantly by a special-purpose implementation, using current state-of-the-art methods such as \cite{caramel-13}, and by choosing convenient parameters, such as finite fields where particularly efficient arithmetic is possible. We concentrate mostly on the computation of the equations of the trace zero subgroup, the factor base, and the relation generation. In particular, we did not implement any filtering (%preprocessing of the matrix, see Remark \ref{rmk:la}),
except for not allowing duplicate relations), we did not implement the double large prime variation, and our implementation is not parallelized.

\subsection{Explicit equations for $n=3$}

When $n=3$, the trace zero variety has dimension 2. Therefore, the index calculus attack on $T_3$ is not more efficient than generic (square root) attacks on $T_3$. Since $n=3$ is the case where all equations are small enough to be written down explicitly,
we present them nevertheless, together with some experimental data that allows us to make predictions on the feasibility of this attack for different values of $q$.

For simplicity, we assume that $3 \mid q-1$ and write $\F_{q^3} = \Fq[\zeta]/(\zeta^3-\mu)$ as a Kummer extension of $\Fq$ with basis $1,\zeta,\zeta^2$. For cases where this is not possible, one may use a normal basis, which gives similar equations. We also assume that $\Fq$ does not have characteristic 2 or 3 and that $E$ is given by an equation in short Weierstra\ss\ form
$$ E : y^2 = x^3 + Ax + B. $$
Our approach also works when $\Fq$ has characteristic 2 or 3, but in this case the definition of the Semaev polynomial and all equations given below must be adjusted (see Remark \ref{rmk:char2ic}).

The third Semaev polynomial is
$$ f_3(z_1,z_2,z_3) = (z_1 - z_2)^2z_3^2 - 2((z_1+z_2)(z_1z_2 + A) + 2B)z_3 + (z_1z_2-A)^2 - 4B(z_1+z_2), $$
and the Weil restriction of $f_3(x,x^q,x^{q^2})$ is
\begin{equation*} %\label{eqn:naumannic}
\begin{array}{rcl}
  \tilde{f}_3(x_0,x_1,x_2) & = & -3x_0^4 - 12 \mu^2 x_0 x_2^3 - 12 \mu x_0 x_1^3 + 18 \mu x_0^2 x_1 x_2\\
  & &  + 9 \mu^2 x_1^2 x_2^2 - 6Ax_0^2 + 6A \mu x_1 x_2 - 12Bx_0 + A^2.
\end{array}
\end{equation*}
We write points of $T_3$ as tuples $(X_0,X_1,X_2)$ that satisfy $\tilde{f}_3(X_0,X_1,X_2) = 0$. For the factor base, we choose those points with $X_0 = 0$, namely
$$ \fb = \{ (0,X_1,X_2) \in T_3 \}.$$
These are precisely the points in $T_3$ that satisfy
\begin{equation} \label{eqn:fb3}
 \tilde{f}_3(0,X_1,X_2) = 9 \mu^2 X_1^2 X_2^2 + 6A \mu X_1 X_2 + A^2 = (3 \mu X_1 X_2 + A)^2 = 0.
\end{equation}
If $A = 0$, then this is equivalent to
$$ X_1 X_2 = 0, $$
and it is particularly easy to enumerate the factor base: One simply checks which $x$-coordinates $(0,X_1,0)$ and $(0,0,X_2)$, for $X_1,X_2 \in \Fq$, give points in $T_3$. If, on the other hand, $A \ne 0$, then every solution of (\ref{eqn:fb3}) satisfies $X_1 \ne 0$, and moreover (\ref{eqn:fb3}) is equivalent to
\begin{equation}\label{eqn:paramfb}
 X_2 = -\frac{A}{3\mu X_1}. 
\end{equation}
In this case, it is also fairly easy to enumerate the factor base: For every $X_1 \in \F_q^{*}$, one computes $X_2$ according to (\ref{eqn:paramfb}) and checks whether this yields a point of $T_3$.

%{\bf actually in this case, it turns out that for all x, the corresponding y-coordinate is in Fq3. but i have no clue why this happens. i guess i should mention it only if i figure out why.}

Now we need to find relations of the form
$$ R = P_0  +  P_1, $$
where $R$ with $x$-coordinate $U = U_0 + U_1\zeta + U_2\zeta^2$ is given and $P_1,P_2$ are in $\fb$. We denote by $x_{P_0} = x_{01}\zeta+x_{02}\zeta^2$ the indeterminates representing the $x$-coordinate of $P_0$ and by $x_{P_1} = x_{11}\zeta+x_{12}\zeta^2$ those representing the $x$-coordinate of $P_1$. Then we have to solve
$$ f_3(x_{P_0},x_{P_1},U) = 0, $$
or equivalently, its Weil restriction.
Assuming that $A \ne 0$, which is the general case,  from (\ref{eqn:paramfb}) we get
$$ x_{02} = -\frac{A}{3\mu x_{01}} \quad \text{and} \quad x_{12} = -\frac{A}{\/3\mu x_{11}}. $$
Plugging this into the above system and multiplying the first two equations by $27\mu x_{01}^2x_{11}^2$ and the third equation by $81\mu^2 x_{01}^2x_{11}^2$ allows us to eliminate the two indeterminates $x_{02}$ and $x_{12}$ from the system that describes a relation. We obtain
\begin{equation}\label{sys:rel3}\small
\begin{array}{rcl}
 0 & = & 36 {x_{01}}^{3}x_{11}{\mu}^{2}AU_{1}U_{2}+36 x_{01}{x_{{11}}}^{3}{\mu}^{2}AU_{1}U_{2}-72 {x_{01}}^{2}{x_{11}}^{2}{\mu}^{2}AU_{1}U_{2}\\
&&-12 x_{01}x_{11}{A}^{2}U_{0}U_{2}\mu+54 {x_{01}}^{4}{x_{11}}^{2}{\mu}^{2}U_{0}U_{1}+54 {x_{{01}}}^{2}{x_{11}}^{4}{\mu}^{2}U_{0}U_{1}\\
&&-18 {x_{01}}^{3}{x_{11}}^{2}{\mu}^{2}AU_{2}-18 {x_{01}}^{2}{x_{11}}^{3}{\mu}^{2}AU_{2}-108 {x_{01}}^{3}{x_{11}}^{3}{\mu}^{2}U_{0}U_{{1}}\\
&&+18 x_{01}{x_{11}}^{4}{\mu}^{2}AU_{2}+18 {x_{01}}^{4}x_{11}{\mu}^{2}AU_{2}+6 {x_{01}}^{2}{A}^{2}U_{0}U_{2}\mu+6 {x_{11}}^{2}{A}^{2}U_{0}U_{2}\mu\\
&&-108 {x_{01}}^{2}{x_{{11}}}^{2}BU_{0}\mu-36 {x_{01}}^{2}{x_{11}}^{2}A{U_{0}}^{2}\mu+6 {x_{01}}^{2}x_{11}{A}^{2}U_{1}\mu+6 x_{01}{x_{11}}^
{2}{A}^{2}U_{1}\mu\\
&&+18 {x_{01}}^{3}x_{11}A{U_{0}}^{2}\mu-6 x_{01}x_{11}{A}^{2}{U_{1}}^{2}\mu+18 x_{01}{x_{11}}^{3}A{U_{0}}^{2}\mu+3 {x_{11}}^{4}{A}^{2}\mu\\
&&+3 {x_{01}}^{4}{A}^{2}\mu-54 {x_{01}}^{4}{x_{11}}^{3}{\mu}^{2}U_{0}-54 {x_{{01}}}^{3}{x_{11}}^{4}{\mu}^{2}U_{0}-54 {x_{01}}^{3}{x_{11}}^{
3}{\mu}^{3}{U_{2}}^{2}\\
&&+27 {x_{01}}^{2}{x_{11}}^{4}{\mu}^{3}{U_{2}}^{2}+27 {x_{01}}^{4}{x_{11}}^{2}{\mu}^{3}{U_{2}}^{2}+18 x_{01}{x_{11}}^{3}{A}^{2}\mu+18 {x_{01}}^{3}x_{11}{A}^{2}\mu\\
&&+39 {x_{01}}^{2}{x_{11}}^{2}{A}^{2}\mu+3 {x_{11}}^{2}{A}^{2}{U_{1}}^{2}\mu+3 {x_{01}}^{2}{A}^{2}{U_{1}}^{2}\mu-6 {x_{01}}^{3}{A}^{2}U_{1}\mu\\
&&-6 {x_{11}}^{3}{A}^{2}U_{1}\mu+2
 x_{01}{A}^{3}U_{0}+2 x_{11}{A}^{3}U_{0}\\
0 & = & -72 {x_{01}}^{2}{x_{11}}^{2}AU_{0}U_{1}\mu+36 x_{01}{x_{11}}^{3}AU_{0}U_{1}\mu-12 x_{01}x_{11}{A}^{2}U_{1}U_{{2}}\mu\\
&&+36 {x_{01}}^{3}x_{11}AU_{0}U_{1}\mu-6 x_{01}x_{{11}}{A}^{3}+3 {x_{01}}^{2}{A}^{2}{U_{0}}^{2}+2 x_{11}{A}^{3}U_{1}+2 x_{01}{A}^{3}U_{1}\\
&&+3 {x_{11}}^{2}{A}^{2}{U_{0}}^{2}-54 {x_{01}}^{3}{x_{11}}^{4}{\mu}^{2}U_{1}+27 {x_{{01}}}^{2}{x_{11}}^{4}{\mu}^{2}{U_{1}}^{2}+27 {x_{01}}^{4}{x_{{11}}}^{2}{\mu}^{2}{U_{1}}^{2}\\
&&-54 {x_{01}}^{3}{x_{11}}^{3}{\mu}^{2}{U_{1}}^{2}-54 {x_{01}}^{4}{x_{11}}^{3}{\mu}^{2}U_{1}-6 {x_{11}}^{3}{A}^{2}U_{2}\mu-6 {x_{01}}^{3}{A}^{2}U_{2}\mu\\
&&-108 {x_{01}}^{2}{x_{11}}^{3}B\mu-108 {x_{01}}^{3}{x_{{11}}}^{2}B\mu-2 {x_{11}}^{2}{A}^{3}+27 {x_{01}}^{4}{x_{11}}^{4}{\mu}^{2}-2 {x_{01}}^{2}{A}^{3}\\
&&-6 x_{01}x_{11}{A}^{2}{U_{{0}}}^{2}+54 {x_{01}}^{2}{x_{11}}^{4}{\mu}^{2}U_{0}U_{2}+54 {x_{01}}^{4}{x_{11}}^{2}{\mu}^{2}U_{0}U_{2}\\
&&-108 {x_{01}}^{3}{x_{11}}^{3}{\mu}^{2}U_{0}U_{2}-36 {x_{01}}^{2}{x_{{11}}}^{2}{\mu}^{2}A{U_{2}}^{2}+18 x_{01}{x_{11}}^{3}{\mu}^{2}A{U_{2}}^{2}\\
&&+18 {x_{01}}^{3}x_{11}{\mu}^{2}A{U_{2}}^{2}-18 {x_{01}}^{3}{x_{11}}^{2}AU_{0}\mu-18 {x_{01}}^{2}{x_{11}}^{3}AU_{0}\mu+6 x_{01}{x_{11}}^{2}{A}^{2}U_{2}\mu\\
&&+6 {x_{{01}}}^{2}x_{11}{A}^{2}U_{2}\mu-108 {x_{01}}^{2}{x_{11}}^{2}BU_{1}\mu+18 x_{01}{x_{11}}^{4}AU_{0}\mu+18 {x_{01}}^{
4}x_{11}AU_{0}\mu\\
&&+6 {x_{01}}^{2}{A}^{2}U_{1}U_{2}\mu+6 {x_{11}}^{2}{A}^{2}U_{1}U_{2}\mu\\
0 & = & -216 {x_{01}}^{2}{x_{11}}^{2}AU_{0}U_{2}{\mu}^{2}+108 x_{{01}}{x_{11}}^{3}AU_{0}U_{2}{\mu}^{2}+108 {x_{01}}^{3}x_{{11}}AU_{0}U_{2}{\mu}^{2}\\
&&+18 {x_{11}}^{2}{A}^{2}U_{0}U_{1}\mu+18 {x_{01}}^{2}{A}^{2}U_{0}U_{1}\mu+18 {x_{01}}^{2}x_{
{11}}{A}^{2}U_{0}\mu+108 x_{01}{x_{11}}^{2}BA\mu\\
&&+18 x_{{01}}{x_{11}}^{2}{A}^{2}U_{0}\mu+108 {x_{01}}^{2}x_{11}BA\mu-36 x_{01}x_{11}{A}^{2}U_{0}U_{1}\mu-162 {x_{01}}^{4}{x_
{11}}^{3}{\mu}^{3}U_{2}\\
&&-162 {x_{01}}^{3}{x_{11}}^{4}{\mu}^{3}U_{2}+9 {x_{11}}^{2}{A}^{2}{U_{2}}^{2}{\mu}^{2}+9 {x_{{01}}}^{2}{A}^{2}{U_{2}}^{2}{\mu}^{2}-162 {x_{01}}^{3}{x_{11}}^{3}A{\mu}^{2}\\
&&+81 {x_{01}}^{2}{x_{11}}^{4}{U_{0}}^{2}{\mu}^{2}-54 {x_{01}}^{2}{x_{11}}^{4}A{\mu}^{2}+81 {x_{01}}^{4}{x_{{1
1}}}^{2}{U_{0}}^{2}{\mu}^{2}-162 {x_{01}}^{3}{x_{11}}^{3}{U_{0}}^{2}{\mu}^{2}\\
&&-54 {x_{01}}^{4}{x_{11}}^{2}A{\mu}^{2}+{A}^{4}-324 {x_{01}}^{2}{x_{11}}^{2}BU_{2}{\mu}^{2}+54 {x_{01}}^{4}x_{11}AU_{1}{\mu}^{2}\\
&&+54 {x_{01}}^{3}x_{11}A{U_{1}}^{2}{\mu}^{2}-18 x_{01}x_{11}{A}^{2}{U_{2}}^{2}{\mu}^{2}+54 x_{01}{x_{11}}^{4}AU_{1}{\mu}^{2}+54 x_{01}{x_{11}}^{3}A{U_{1}}^{2}{\mu}^{2}\\
&&-54 {x_{01}}^{3}{x_{11}}^{2}AU_{1}{\mu}^{2}-54 {x_{01}}^{2}{x_{11}}^{3}AU_{1}{\mu}^{2}-108 {x_{{01}}}^{2}{x_{11}}^{2}A{U_{1}}^{2}{\mu}^{2}\\
&&+162 {x_{01}}^{4}{x_{11}}^{2}{\mu}^{3}U_{1}U_{2}-324 {x_{01}}^{3}{x_{11}}^{3}{\mu}^{3}U_{1}U_{2}+162 {x_{01}}^{2}{x_{11}}^{4}{\mu}^{3}U
_{1}U_{2}\\
&&-18 {x_{11}}^{3}{A}^{2}U_{0}\mu+6 x_{11}{A}^{3}U_{2}\mu-18 {x_{01}}^{3}{A}^{2}U_{0}\mu+6 x_{01}{A}^{3}U_{{2}}\mu.
\end{array}
\end{equation}
This system only involves the two indeterminates $x_{01},x_{11}$. All equations have degree 4 in both $x_{01}$ and $x_{11}$. The first and third equations have total degree 7, and the second equation has total degree 8. We have computed that the system (\ref{sys:rel3}) has regularity 14 for almost all points $R$ (and regularity 12 or 13 for some special choices of $R$). This means that the highest degree of all polynomials appearing during the Gr\"obner basis computation is at most 14. This moderate number suggests that the Gr\"obner basis computation is not very costly, and our experiments (see below) show that this is indeed true.

For a given $x$-coordinate $U$ of a point $R \in T_n$, the $\Fq$-solutions $(X_{01},X_{11})$ of the above system with $X_{01},X_{11} \ne 0$ give candidates for $x$-coordinates 
$$ X_0 = X_{01}\zeta - \frac{A}{3\mu X_{01}}\zeta^2 \quad \text{and} \quad X_1 = X_{11}\zeta - \frac{A}{3 \mu X_{11}}\zeta^2 $$
of the points $P_0,P_1$ in the relation.

\begin{example}
 We give a toy example. Let $q = 2^{12}-3, \F_{q^3} = \Fq/(\zeta^3-2)$, and $E : y^2 = x^3 + x + 21$. Then $T_3$ has order 16715869, which is a 24-bit prime, and we take
 $$ P = 3961 + 199\zeta + 4028\zeta^2 $$
as a generator. We choose a random
$$ Q = 3342 + 3020\zeta + 4031\zeta^2 , $$
of which we wish to compute the discrete logarithm. The elements of the factor base satisfy
$$ 6x_{i1}x_{i2} + 1 = 0, \quad i = 0,1, $$
and we compute that there are exactly 4002 such points. Now we choose random $\alpha = 4297188$ and $\beta = 10382682$, which gives $U = 2960 + 1129\zeta + 1917\zeta^2$, and we solve the system
\begin{equation*}\small
\begin{array}{rcl}
0 & = &    439 x_{01}^4 x_{11}^3 + 1215 x_{01}^4 x_{11}^2 + 2556 x_{01}^4 x_{11} + 2274 x_{01}^4 + 439 x_{01}^3 x_{11}^4 + 1663 x_{01}^3 x_{11}^3\\
&& + 1537 x_{01}^3 x_{11}^2
        + 3403 x_{01}^3 x_{11} + 2023 x_{01}^3 + 1215 x_{01}^2 x_{11}^4 + 1537 x_{01}^2 x_{11}^3 + 1961 x_{01}^2 x_{11}^2\\
        && + 2070 x_{01}^2 x_{11} + 
        2326 x_{01}^2 + 2556 x_{01} x_{11}^4 + 3403 x_{01} x_{11}^3 + 2070 x_{01} x_{11}^2 + 3534 x_{01} x_{11}\\
        && + 716 x_{01} + 2274 x_{11}^4 + 2023 x_{11}^3 + 
        2326 x_{11}^2 + 716 x_{11}\\
0 & = &    2 x_{01}^4 x_{11}^4 + 3670 x_{01}^4 x_{11}^3 + 938 x_{01}^4 x_{11}^2 + 609 x_{01}^4 x_{11} + 3670 x_{01}^3 x_{11}^4 + 2217 x_{01}^3 x_{11}^3\\
&& + 
        3400 x_{01}^3 x_{11}^2 + 405 x_{01}^3 x_{11} + 3667 x_{01}^3 + 938 x_{01}^2 x_{11}^4 + 3400 x_{01}^2 x_{11}^3 + 2586 x_{01}^2 x_{11}^2\\
        && + 
        426 x_{01}^2 x_{11} + 94 x_{01}^2 + 609 x_{01} x_{11}^4 + 405 x_{01} x_{11}^3 + 426 x_{01} x_{11}^2 + 115 x_{01} x_{11}\\
        && + 345 x_{01} + 3667 x_{11}^3 + 
        94 x_{11}^2 + 345 x_{11}\\
0 & = &    518 x_{01}^4 x_{11}^3 + 1692 x_{01}^4 x_{11}^2 + 2117 x_{01}^4 x_{11} + 518 x_{01}^3 x_{11}^4 + 2070 x_{01}^3 x_{11}^3 + 1976 x_{01}^3 x_{11}^2\\
&& + 
        1677 x_{01}^3 x_{11} + 1945 x_{01}^3 + 1692 x_{01}^2 x_{11}^4 + 1976 x_{01}^2 x_{11}^3 + 3431 x_{01}^2 x_{11}^2 + 2162 x_{01}^2 x_{11}\\
        && + 1057 x_{01}^2 
        + 2117 x_{01} x_{11}^4 + 1677 x_{01} x_{11}^3 + 2162 x_{01} x_{11}^2 + 1979 x_{01} x_{11} + 71 x_{01}\\
        && + 1945 x_{11}^3 + 1057 x_{11}^2 + 71 x_{11} +         3474.
\end{array}        
\end{equation*}
We get $X_{01} = 1770, X_{11} = 1515$, and from these we compute $X_{02} = 338, X_{12} = 3029$, which gives a relation
$$ P_0  +  P_1 = R $$
for some choice of $y$-coordinates. After collecting 4002 more such relations and solving the linear system, we obtain $\log_P Q = 419$.
\end{example}

%\begin{remark}\label{rmk:gb3}
% During the relation collection phase, a system of similar shape must be solved many times. This is typical for such an index calculus attack, and different techniques exist to make relation generation more efficient. One idea is to compute a parametric Gr\"obner basis \cite{weispfenning-92,montes-wibmer-10}, where $U_0,\ldots,U_{n-1}$ are treated as parameters. The idea is to carry out a Gr\"obner basis computation only once, and afterwards, to immediately obtain a Gr\"obner basis of of the system where numbers are plugged in for the parameters by plugging the numbers into the parametric Gr\"obner basis. However, the cost of such computations is very large, and we were not able to compute a parametric Gr\"obner basis even in this small case where $n=3$. A more efficient approach, called Gr\"obner trace algorithm, is due to Traverso \cite{traverso-89} and was implemented by Joux and Vitse \cite{joux-vitse-11}. Their variant of the F4 Gr\"obner basis algorithm is particularly suitable for computing Gr\"obner bases of a number of systems of the same shape, since it can reuse information from the first computation for all subsequent ones. However, since we were readily able to solve our system for $n=3$, and since this is not the interesting case in terms of complexity, we did not try to use their algorithm.
%\end{remark}

Finally, we present implementation results for fields of different size in Table \ref{table:indexcalc3}. For primes $q$ of $10,12,14,16,18,20,30,40,50,60,70$, and $80$ bits, we chose the smallest possible value $\mu$, and we chose curves $E$, given by the coefficients $A,B$, that yield cyclic trace zero subgroups $T_3$ of prime order. Where we were able to compute it, we list the exact size of the factor base. In all cases, it is close to $q - q^{1/2}$. We also list the number of points $R$ we had to try in order to find $|\fb|+1$ distinct relations.

Times are given in seconds, and numbers in normal font stand for computations that we were able to perform, while numbers in bold represent expected times, extrapolated from timings we were able to obtain. For example, when we are able to compute one relation, this allows us to predict the time it would take to collect $q$ relations (experimentally this requires solving about $2q$ polynomial systems). Where we were not able to carry out a computation or make a prediction, we write ``--''.

For all field sizes, we were able to solve the system at least a few times. For comparison, we give the time taken to compute a lexicographic Gr\"obner basis of the straightforward system consisting of 5 equations in 4 indeterminates (``large system''), as well as the time taken to compute a lexicographic Gr\"obner basis of system (\ref{sys:rel3}) consisting of 3 equations in 2 indeterminates (``small system''). This shows that this little trick to simplify the system saves a considerable amount of time in practice. Therefore, in the following, we work with the small system.

Next we list in the table the average time taken to solve the small system once. This includes computing the lexicographic Gr\"obner basis, factoring a univariate polynomial (of degree 6 in our experiments), which gives the value(s) of one indeterminate, and computing the corresponding value(s) of the other indeterminate. For the Gr\"obner basis computation, we use Magma's {\tt GroebnerBasis()}, which computes a degree reverse lexicographic Gr\"obner basis using Faug\`ere's F4 algorithm \cite{f4} and subsequently a lexicographic Gr\"obner basis using the FGLM algorithm \cite{fglm}.

Finally, we give the actual or extrapolated times for the full execution of the different steps of our algorithm. First we give the time to enumerate the factor base, then the time to collect $|\fb|+1$ relations, and then the time to solve the linear system, using the sparse linear algebra routine {\tt ModularSolution(Lanczos:=true)} of Magma, which is an implementation of Lanczos' algorithm. We also give the total time to compute one discrete logarithm with our algorithm.

\begin{table}\small
\caption{Index calculus algorithm for $n=3$, timings in seconds}
\label{table:indexcalc3}
\begin{tabular}{l|l|l|l|l|l|l}
\hline\noalign{\smallskip}
$\log_2 |T_3|$  & 20 & 24 & 28 & 32 & 36 & 40 \\
\noalign{\smallskip}\hline\noalign{\smallskip}
$q$ & $2^{10}-3$ & $2^{12}-3$ & $2^{14}-3$ & $2^{16}-15$ & $2^{18}-93$ & $2^{20}-3$ \\
$\mu$ & 5 & 2 & 2 & 2 & 2 & 2 \\
$A$ & 2 & 1 & 1& 1 & 1 & 1 \\
$B$ & 0 & 21 & 11 & 5 & 10 & 25 \\
$|\fb|$                         & 900     & 4002    & 16380   & 65388   & 261822  & 1045962 \\
number of $R$'s tried           & 2208    & 8263    & 32828   & 130533  & 522935  & 2091965 \\
time for GB of large system     & 0.01773 & 0.01698 & 0.01705 & 0.01792 & 0.01686 & 0.01703 \\
time for GB of small system     & 0.00102 & 0.00169 & 0.00167 & 0.00124 & 0.00146 & 0.00135 \\
time to solve small system      & 0.00115 & 0.00180 & 0.00173 & 0.00134 & 0.00159 & 0.00136 \\
time to enumerate $\fb$         & 0.07    & 0.28    & 1.15    & 5.24   & 23.59 & 104.86 \\
time to collect relations & 3.52    & 13.53   & 49.71   & 197.17  & 803.95  &  2845.01 \\
time linear algebra             & 0.01    & 0.30    & 5.22    & 108.29  & 129.69   & -- \\
total time                      & 3.60    & 14.25   & 56.08   & 310.70  & 957.23 & -- \\
\noalign{\smallskip}\hline\noalign{\smallskip}
$\log_2 |T_3|$  & 60 & 80 & 100 & 120 & 140 & 160 \\
\noalign{\smallskip}\hline\noalign{\smallskip}
$q$ & $2^{30}-105$ & $2^{40}-87$ & $2^{50}-51$ & $2^{60}-93$ & $2^{70}-267$ & $2^{79}-67$ \\
$\mu$ & 2 & 2 & 2 & 2 & 5 & 3 \\
$A$ & 1 & 1 & 1& 1 & 1 & 1 \\
$B$ & 24 & 49 & 40 & 193 & 15 & 368 \\
$|\fb|$                     & $\mathbf{2^{30}}$ & $\mathbf{2^{40}}$ & $\mathbf{2^{50}}$ & $\mathbf{2^{60}}$ & $\mathbf{2^{70}}$ & $\mathbf{2^{79}}$ \\
number of $R$'s tried       & $\mathbf{2^{31}}$ & $\mathbf{2^{41}}$ & $\mathbf{2^{51}}$ & $\mathbf{2^{61}}$ & $\mathbf{2^{71}}$ & $\mathbf{2^{80}}$ \\
time for GB of large system & 0.02683 & 0.12645 & 0.12817 & 0.13431 & 0.15000 & 0.14102 \\
time for GB of small system & 0.00146 & 0.00231 & 0.00244 & 0.00249 & 0.00304 & 0.00262 \\
time to solve small system  & 0.00171 & 0.00291 & 0.00342 & 0.00351 & 0.00467 & 0.00442 \\
time to enumerate $\fb$     & $\mathbf{2^{17.2}}$ & $\mathbf{2^{28.3}}$ & $\mathbf{2^{38.5}}$ & $\mathbf{2^{48.7}}$ & $\mathbf{2^{59.4}}$ & $\mathbf{2^{68.4}}$ \\
time to collect relations & $\mathbf{2^{21.8}}$ & $\mathbf{2^{32.5}}$ & $\mathbf{2^{42.8}}$ & $\mathbf{2^{52.8}}$ & $\mathbf{2^{63.2}}$ & $\mathbf{2^{72.1}}$ \\
%time linear algebra         & -- & -- & -- & -- & -- & --\\
%total time                  & -- & -- & -- & -- & -- & --\\
\noalign{\smallskip}\hline
\end{tabular}
\end{table}

We see that the largest trace zero subgroup where we can compute a full discrete logarithm with our prototype implementation has 36-bit size. The attack takes approximately 15 minutes. For the 40-bit trace zero subgroup, we can compute sufficiently many relations in about 47 minutes, but we are not able to solve the linear system of size about $2^{20} \times 2^{20}$ in Magma. A specialized implementation presented in \cite{caramel-13,jeljeli-12,jeljeli-14} solves a linear system of size about $2^{22} \times 2^{22}$ in less than 5 days using a sophisticated implementation of Lanczos' algorithm, running on a high performance computer. This means that our attack is certainly feasible for a 40-bit trace zero subgroup. However, we can do much better by rebalancing the cost of relation collection and linear algebra. 

Let us consider e.g. the group $T_3$ of 60 bits, with $q \approx 2^{30}$, as given in Table \ref{table:indexcalc3}. We rebalance the complexity with a relatively straightforward approach. Using a factor base of $q^r = 2^{30r}$ elements, where  $0 < r < 1$, the probability of finding a relation becomes $q^{2r-2}/2$. Hence in order to find $q^r$ relations, we need to solve $2q^{2-r} = 2^{61-30r}$ systems. Since we know that solving a linear system of size $2^{22} \times 2^{22}$ is possible, we set $q^r = 2^{22}$ and get $r = 0.73$. This means that we would have to collect $2^{39}$ relations, which would take $2^{39.8}$ seconds or about 30 years. Assuming that solving a linear system of size $2^{23} \times 2^{23}$ is possible, we would need about 15 years to collect relations, etc. We stress that these predictions correspond to the time required by our simple implementation. With an optimized and parallel implementation of the relation collection step (notice that the relation search can trivially be parallelized), it would become faster by a considerable factor. Hence we conclude that with an optimized implementation, computing a discrete logarithm in a 60-bit trace zero subgroup with this index calculus algorithm is feasible.

\subsection{Explicit equations for $n=5$} \label{sec:explicit5ic}

We proceed simliarly for $n = 5$, but we do not write down the equations in this case because they are too large. We assume that $5 \mid q-1$ and write $\F_{q^5} = \Fq(\zeta)/(\zeta^5-\mu)$. Then $1,\zeta,\zeta^2,\zeta^3,\zeta^4$ is a basis of $\F_{q^5}|\Fq$, which we use for Weil restriction.

The fifth Semaev polynomial $f_5$ has total degree 32. The same is true for $\tilde{f_5}(x_0,\ldots,x_4)$, which we use as an equation for $T_5$. The factor base is
$$ \fb = \{ (0,0,0,X_3,X_4) \in T_5\}, $$
and all its elements satisfy the equation
\begin{equation}\label{eqn:fb}
\tilde{f_5}(0,0,0,x_3,x_4) = 0.\end{equation}
It has total degree 32 and degree 30 in each $x_3$ and $x_4$. Although this polynomial does not have such a simple shape as the corresponding one for $n=3$, it is still easy to enumerate the factor base: For every $X_3 \in \Fq$, solve $\tilde{f_5}(0,0,0,X_3,x_4) = 0$ for $x_4$ in $\Fq$.

Following an idea of Joux and Vitse \cite{joux-vitse-12} (see Remark \ref{rmk:jouxvitse}), we look for relations of the form
\begin{equation}\label{eqn:sum}
R = P_0  +  P_1  +  P_2,\end{equation}
where $P_0, P_1, P_2$ are elements of the factor base. We obtain a system of 8 equations in 6 indeterminates: The first 5 equations are the Weil restriction of $f_4(x_{P_0},x_{P_1},x_{P_2},U)$ and correspond to (\ref{eqn:sum}). They have total degree 12 and degree 4 in each indeterminate. The last 3 equations correspond to the condition that the points belong to the factor base and are of the form (\ref{eqn:fb}).
For a given $U$, we solve this system in order to obtain possible relations.
However, the system is too large to be solved with Magma. Even over the relatively small field $\F_{1021}$, our computation did not finish after several weeks of computation and using more than 300 GB of memory.

Hence we use a hybrid approach along the lines of \cite{yang-chen-courtois-04,bettale-faugere-perret-08}. This allows us to find some relations, but it is not fast enough for an attack of realistic cryptographic size. Nevertheless, we give some experimental results, timings, and extrapolations. The hybrid method is often used where a direct Gr\"obner basis computation is too costly, since it is a trade-off between exhaustive search and Gr\"obner basis techniques. The main idea is to choose fixed values for a small number of variables and to solve the system in the remaining indeterminates. In order to find all solutions of the system, all choices for the fixed variables have to be tried. Therefore, this requires computing many Gr\"obner bases of smaller systems instead of computing one Gr\"obner basis of a large system.

In our case, it is enough to choose one fixed value in order to solve the system readily. %We start from the system that describes a relation with 3 factor base elements (i.e.\ following the approach of \cite{joux-vitse-12}) and 
We fix $x_{03} = X_{03} \in \Fq$ and use the factor base equation $\tilde{f}_5(0,0,0,X_{03},x_{04}) = 0$ to determine possible values of $x_{04}$. Although this equation has degree 30 in $x_{04}$, there are usually only very few solutions, most frequently 1, 2, or 3. In every case where $x_{04} = X_{04}$ gives a point in the factor base, we plug $x_{03} = X_{03}$ and $x_{04} = X_{04}$ into the system to obtain a new system of 7 equations in the 4 indeterminates $x_{13},x_{14},x_{23},x_{24}$. The first five equations each have total degree 8 and degree 4 in every indeterminate. By trying all $X_{03} \in \Fq$, we find out whether $R$ decomposes over the factor base. %{\bf if i can compute it, might comment here on the regularity of this system. however right now it looks to me like it can't be done: while a GB of the original system is easy, after homogenizing all the equations i end up with something where a GB computation is difficult.}

We give some timings and extrapolations in Table \ref{table:indexcalc5}. As before, numbers in normal font are times we measured, and numbers in bold are predictions. After giving the parameters of the fields and curves we used, we indicate the number of points $R$ which we tried to decompose (we expect $6q$), the total number of polynomial systems to be solved for this (we expect $6q^2$), the time for the solution of one system (this is equal to the time for computing a Gr\"obner basis, since the rest of the computation to solve the system is negligible), the time to enumerate the factor base, the time to collect about $q$ relations, the time for the linear algebra step, and the time for the total attack.

The numbers show that we are able to compute a discrete logarithm in the 27-bit group $T_5$ in about 2 days and that a discrete logarithm in the 32-bit, 36-bit, and 40-bit groups $T_5$ can be computed in about 10, 44, and 165 days, respectively. Clearly, this approach is far from feasible for any group of cryptographic size.

\begin{table}\small
\caption{Index calculus algorithm for $n=5$, timings in seconds}
\label{table:indexcalc5}
\begin{tabular}{l|l|l|l|l|l|l}
\hline\noalign{\smallskip}
$\log_2 |T_5|$  & 20 & 22 & 27 & 32 & 36 & 40 \\
\noalign{\smallskip}\hline\noalign{\smallskip}
$q$ & $2^{5}-1$ & $2^{6}-23$  & $2^{7}-27$ & $2^{8}-15$ & $2^{9}-21$ & $2^{10}-3$ \\
$\mu$ & 2 & 2 & 2 & 3 & 2 & 2 \\
$A$ & 1 & 1 & 1 & 1 & 1 & 1 \\
$B$ & 16 & 3 & 3 & 13 & 18 & 1 \\
$|\fb|$ & 40 & 70 & 110 & 230 & 520 & 970 \\
number of $R$'s tried & 886 & 884 & 2424 & {\bf 5784} & {\bf 11784} & {\bf 24528} \\
number of systems solved & 17719 & 30934 &  244824& {\bf 1393944} & {\bf 5785944} & {\bf 25043088} \\
time for GB of one system & 1.30 & 1.31 & 1.28 & 1.21 & 1.22 & 1.32 \\
%time to solve one system & 1.42 & 1.41 &  &  &  &  \\
%average time to find one relation & 559.186 & 573.826 &  &  &  &  \\
%average time per $R$ & 26.12 & 54.99 & 70.58 & 142.91 & 324.66 & 615.32 \\
time to enumerate $\fb$ & 0.02 & 0.04 & 0.07 & 0.18 & 0.43 & 0.89 \\
time to collect relations & 25004 & 38219 & 171085 & {\bf 821328} & {\bf 3818016} & {\bf 15084720} \\
time linear algebra & 0.01 & 0.01 & 0.01 & 0.01 & 0.01 & 0.01\\
total time & 25164 & 43618 & 171085 & {\bf 821328} & {\bf 3818016} & {\bf 15084720} \\
\noalign{\smallskip}\hline\noalign{\smallskip}
$\log_2 |T_5|$  & 60 & 80 & 100 & 120 & 140 & 160 \\
\noalign{\smallskip}\hline\noalign{\smallskip}
$q$ & $2^{15}-157$ & $2^{20}-5$ & $2^{25}-61$ & $2^{30}-173$ & $2^{35}-547$ & $2^{40}-195$ \\
$\mu$ & 3 & 2 & 2 & 2 & 5 & 2 \\
$A$ & 1 & 1& 1 & 1 & 1 & 1 \\
$B$ & 7 & 10 & 17 & 5 & 3 & 12 \\
$|\fb|$ & 32600 & 1051440 & $\mathbf{2^{25}}$ &$\mathbf{2^{30}}$ & $\mathbf{2^{35}}$ & $\mathbf{2^{40}}$ \\
number of $R$'s tried & $\mathbf{2^{20}}$ & $\mathbf{2^{25}}$ & $\mathbf{2^{30}}$ & $\mathbf{2^{35}}$ & $\mathbf{2^{40}}$ & $\mathbf{2^{45}}$ \\
number of systems solved & $\mathbf{2^{35}}$ & $\mathbf{2^{45}}$ & $\mathbf{2^{55}}$ & $\mathbf{2^{65}}$ & $\mathbf{2^{75}}$ & $\mathbf{2^{85}}$ \\
time for GB of one system & 1.34 & 1.33 & 7.09 & 6.93 & 146.16 & 147.89 \\
%time to solve one system &  &  &  &  &  &  \\
%average time to find one relation &  &  &  &  &  &  \\
%average time per $R$ & 20716.42 &  &  &  & &  \\ 
time to enumerate $\fb$ & 38.80 & 1530.91 & $\mathbf{2^{17.1}}$ & $\mathbf{2^{22.9}}$ & $\mathbf{2^{28.7}}$  & $\mathbf{2^{34.0}}$  \\
time to collect relations & $\mathbf{2^{34.3}}$ & $\mathbf{2^{45.4}}$ & $\mathbf{2^{57.8}}$ & $\mathbf{2^{67.7}}$ & $\mathbf{2^{82.2}}$ & $\mathbf{2^{92.2}}$\\
time linear algebra & 89.12 & -- & -- & -- & -- & --\\
total time & $\mathbf{2^{34.3}}$  & $\mathbf{2^{45.4}}$ & $\mathbf{2^{57.8}}$ & $\mathbf{2^{67.7}}$ & $\mathbf{2^{82.2}}$ & $\mathbf{2^{92.2}}$\\
\noalign{\smallskip}\hline
\end{tabular}
\end{table}

We see that it is very costly to find a relation with this approach, for two reasons. Firstly, we are searching for relations that involve only 3 points of the factor base. While the probability that a point decomposes into a sum of 4 points of the factor base is $1/4! = 1/24$, the probability that it decomposes into a sum of 3 points of the factor base is $1/(3!q) = 1/(6q)$ (%using similar heuristics as the complexity analysis,
see Section \ref{sec:complexity}). This means that we expect to have to try about $6q$ points $R$ in order to find one that decomposes. Notice that we can still hope to find enough relations, even though the probability of finding a relation has decreased by a factor $q$ (Joux and Vitse \cite{joux-vitse-12} have shown that such an approach is indeed advantageous in certain situations): Assuming that most distinct unordered $3$-tuples of factor base elements sum to distinct points of $T_5$, this means that about $q^3/6$ points $R \in T_5$ decompose into a sum of 3 factor base elements. This number is much larger than $q$. Therefore, it is a realistic assumption that we find about $q$ relations.

Secondly, every time we wish to check whether a given point $R$ decomposes into a sum of 3 factor base points, we do not have to solve one system, but $O(q)$ systems, namely a small number of systems for every $X_{03} \in \Fq$. In practice, not all $X_{03}$ yield valid $X_{04}$, therefore the number of systems to be solved is actually smaller.

\section{Comparison with other attacks and discussion} \label{sec:comparison}

We now compare the index calculus attack on the DLP in $T_n$ with other known attacks.

\subsection{Pollard--Rho}
Assuming that $T_n$ is cyclic of prime order, the Pollard--Rho Algorithm performs $O(q^{(n-1)/2})$ steps, and each step consists essentially of a point addition and hence has complexity $\tilde{O}(1)$. Comparing this to the complexity of the index calculus algorithm in $q$, which is $\tilde{O}(q^{2-2/(n-1)}),$ we see that the index calculus algorithm has smaller complexity for $n\geq 5$. More precisely, when $n=3$ then Pollard--Rho and index calculus have the same complexity, when $n=5$ the advantage of the index calculus attack comes only from the large prime variation (because without the large prime variation, index calculus has complexity $\tilde{O}(q^2)$), and when $n > 5$, the index calculus method always has lower complexity, independently of the large prime trick. The larger $n$, the larger the advantage of the index calculus algorithm over Pollard--Rho in this analysis. 

However, the Pollard--Rho Algorithm has to perform only an elliptic curve point addition in each step, while the index calculus algorithm has to compute a Gr\"obner basis, which is much more expensive. Even in the case $n=3$, where the system is much more manageable than for larger $n$, we can solve less than a thousand systems per second (cf.\ Table \ref{table:indexcalc3}), whereas elliptic curve point addition can be performed at a rate of 25000 to 150000 per second (depending on the size of the field; we measured this by adding random points of $T_3$ in Magma, an optimized implementation can achieve much better values). For larger values of $n$, the difference becomes much more extreme, since the cost of elliptic curve point addition increases only at the same rate as that of finite field arithmetic in $\Fqn$, whereas the cost of the Gr\"obner basis computation increases considerably. In fact the degree of the equations grows exponentially and the number of equations and variables grows linearly in $n$. 
%This becomes very obvious in our experiments, where we conclude that we cannot solve one single system for $n=5$, and it is visible also 
This is reflected in the large complexity in $n$ of the index calculus algorithm (see Theorem \ref{thm:indexcalccomplexity}).

We conclude that in practice index calculus can be more efficient than Pollard--Rho only for moderate values of $n > 3$ and very large values of $q$. We do not know the precise crossover point.

Notice also that the variant of the index calculus algorithm for $T_5$ that uses the trick of Joux and Vitse and the hybrid approach has complexity $\tilde{O}(q^3)$ in $q$, therefore it is not better than the Pollard--Rho Algorithm for $n=5$. It would be better only for $n > 5$.

\subsection{Index calculus on the whole curve}

The index calculus algorithm of Gaudry may also be used to compute discrete logarithms in $E(\Fqn)$ by working in the $n$-dimensional Weil restriction of $E$ with respect to $\Fqn|\Fq$. This is one of the original applications suggested by Gaudry in \cite{gaudry-09}. From a complexity theoretic point of view, it does not make sense to attack the DLP in $E(\Fqn)$ when one wants to solve a DLP in $T_n$, since the complexity of Gaudry's algorithm in $q$ depends on the dimension of the variety and therefore has complexity $\tilde{O}(q^{2-2/n})$ in $E(\Fqn)$ and complexity $\tilde{O}(q^{2-2/(n-1)})$ in $T_n$.

From a practical point of view, however, the systems one gets when performing index calculus on the whole curve may be more manageable, since they consist only of the Weil restriction of the Semaev polynomial, whereas in our approach, the system contains also the equations of the factor base. Moreover, when working in the whole curve, the Semaev polynomial may easily be symmetrized, which gives a system of smaller degree and with fewer solutions, whereas it is not obvious how to do this in our case. Also, when working in the whole curve, factor base elements may be represented by one $\Fq$-coordinate only, where we need two for the trace zero variety. Therefore, our system has twice as many indeterminates. On the other hand, the advantage of working in the trace zero variety is that relations contain $n-1$ factor base elements, and therefore one uses $f_n$ to describe relations, whereas when working on the whole curve, relations contain $n$ factor base elements, thus one has to use $f_{n+1}$. Summarizing, when working on the whole curve, one has a system of $n$ equations in $n$ indeterminates of total degree $2^{n-1}$. In contrast, when working in the trace zero variety one has a system of $2n-1$ equations in $2n-2$ indeterminates of total degree $(n-1)2^{n-2}$.

Such subtleties are not evident in the original complexity analysis of Gaudry, which is only in $q$ (and $n$ is taken to be constant) and where the Gr\"obner basis computation thus has constant complexity. When performing an analysis similar to the one of Section \ref{sec:complexity} for Gaudry's algorithm on the whole curve, one obtains 
 $$ \tilde{O}\left({n2^{n-1} + 1 \choose n}^{\omega}n!q^{2-2/n}\right), $$
which is smaller in $n$. Therefore, which attack performs better depends on the relation between $q$ and $n$.

%Summarizing, it is difficult to accurately predict the running time of index calculus algorithms, and
In both cases the feasibility of the Gr\"obner basis computation plays an important role in practice.

\subsection{Cover attacks}

Cover attacks, also referred to as transfer attacks, were first proposed by Frey \cite{frey-99} and further studied by many authors, including Galbraith and Smart \cite{galbraith-smart-99}, Gaudry, Hess, and Smart \cite{ghs}, and Diem \cite{diem-ghs}. The aim of such attacks is to transfer the DLP from the algebraic variety one is considering to the Picard group of a curve of larger (but still rather low) genus, where the DLP is then solved using index calculus methods. There exist different constructions, each of them specific to a certain type of curve or variety, and there are constructions for cover attacks on $E(\Fqn)$ and on $T_n$ directly.

For example, combining the results of \cite{diem-ghs} and \cite{diem-kochinke}, it is sometimes possible to map the DLP to the Picard group of a
genus 5 curve (which is usually not hyperelliptic), where it can be solved in $\tilde{O}(q^{4/3})$ . This is better than Gaudry's index calculus in $E(\F_{q^5})$, which has complexity $\tilde{O}(q^{8/5})$, and the index calculus attack on $T_5$, which has complexity $\tilde{O}(q^{3/2})$. However, the index calculus attack on $T_5$ applies to all curves, whereas only a very small proportion of curves is affected by the cover attack.

Diem and Scholten \cite{diem-scholten,diem-scholten-arehcc} propose a cover attack for the trace zero variety directly. It works best for trace zero varieties of genus 2 curves, but it also applies to some trace zero varieties of elliptic curves. Namely, when $g=1$ and $n=5$, the DLP may sometimes be transferred to a curve of genus 4, where it can be solved in $\tilde{O}(q^{4/3})$. Again, this is better than the complexity of the index calculus attack, but it only affects a small number of curves (in fact, in \cite{diem-scholten-arehcc} the authors find only one curve vulnerable to this attack). The same is true for $g=1$ and $n=7$, where the DLP may sometimes be mapped to a curve of genus 8 (in this case the authors cannot find any examples, although they can prove that vulnerable curves exist).

%Joux and Vitse \cite{joux-vitse-12} propose a cover and decomposition attack, which combines a cover attack with index calculus. However, this only applies to elliptic curves over composite degree extension fields, and therefore it does not threaten the trace zero variety.

\section{Conclusions on the hardness of the DLP}\label{sec:securitydiscussion}

We conclude that applying Gaudry's index calculus algorithm for abelian varieties to the trace zero variety, as presented in this paper, yields an attack in $T_n$ that has smaller complexity than generic algorithms whenever $n\geq 5$ when the complexity is measured asymptotically in $q$. Although there sometimes exist cover attacks with even better complexity, the index calculus attack can be applied to trace zero varieties of all elliptic curves, while cover attacks apply only to a small proportion of curves.

Since the DLP in $T_n$ has the same complexity as the DLP in $E(\Fqn)$, we get that the DLP in $E(\Fqn)$ may be attacked in complexity $\tilde{O}(q^{2-2/(n-1)})$ when $E$ is defined over $\Fq$. This is better than all known direct attacks on the DLP in $E(\Fqn)$ for $n\geq 5$.
% The most interesting case in this context is when $n=3$. Here generic attacks on $E(\F_{q^3})$ have complexity $O(q^{3/2})$, Gaudry's index calculus attack applied to $E(\F_{q^3})$ has complexity $\tilde{O}(q^{4/3})$, and our index calculus attack on $T_3$ has complexity $\tilde{O}(q)$. Moreover, we have seen that our algorithm is practical for small $q$ in this case, since the system has a particularly simple shape and can be solved rapidly. This becomes faster than Pollard--Rho for values of $q$ larger than about 60 bits, and asymptotically it yields one of the best attacks on the DLP in $E(\F_{q^3})$. Notice, however, that the Pollard--Rho Algorithm in $T_3$ has the same complexity and is much faster in practice.

For general $n$, we have seen that the complexity of our index calculus attack on $T_n$ depends exponentially on $n$ and that it becomes infeasible for rather small values of $n$. This is due to the fact that the algorithm has to solve many polynomial systems, whose size (i.e.\ number of equations, number of indeterminates, degrees of the equations) depends on $n$, and that a Gr\"obner basis computation quickly becomes unmanageable. In fact, already for $n=5$ we cannot solve the system with standard Gr\"obner basis software. By using some tricks (namely, considering relations that involve one point less, using a hybrid approach), we were nevertheless able to produce relations. However this does not yield a practical attack, since it multiplies the complexity of the relation search by a factor $q^2$.

Specialized Gr\"obner basis techniques in the spirit of \cite{joux-vitse-11,faugere-perret-petit-renault-12,petit-quisquater-12} would be needed in order to efficiently solve the systems that arise in this index calculus attack, and more research needs to be done on this topic in order to make our index calculus attack feasible in practice.

\bibliographystyle{amsalpha}  
\bibliography{lit}

\end{document}